\documentclass[twocolumn,10pt,notitlepage]{article}
\usepackage{geometry} 
\usepackage[sort&compress,numbers,super]{natbib}%
\usepackage{doi,url}
\usepackage{amsmath,amssymb}
\usepackage{graphicx}
\usepackage{times}

\title{Search for domain wall dark matter with atomic clocks on board Global Positioning System satellites}

\author{
Benjamin M. Roberts,$^{1}$ 
Geoffrey Blewitt,$^{1,2}$ Conner Dailey,$^{1}$ 
Mac Murphy,$^{1}$  Maxim Pospelov,$^{3,4}$\\
 Alex Rollings,$^{1}$ Jeff Sherman,$^{5}$ Wyatt Williams,$^{1}$ Andrei Derevianko$^{\ast1}$\\
~\\[-0.25cm]
\small{$^{1}$Department of Physics, University of Nevada, Reno, NV 89557, USA};~\\
\small{$^{2}$Nevada Geodetic Laboratory, Nevada Bureau of Mines and Geology, University of Nevada, Reno, NV 89557, USA};~\\
\small{$^{3}$Department of Physics and Astronomy, University of Victoria, Victoria, BC V8P 1A1, Canada};~\\
\small{$^{4}$Perimeter Institute for Theoretical Physics, Waterloo, ON N2J 2W9, Canada};~\\
\small{$^{5}$National Institute of Standards and Technology, Boulder, CO 80305, USA}.
}

\date{19 September 2017}

\usepackage{hyperref} 
\usepackage{xcolor}
\definecolor{cite}{rgb}{0.,0.,0.5}   
\hypersetup{ 
	linktoc=section,
	colorlinks,
	allcolors={cite},
}

\newcommand{\natabstract}[1]{
\twocolumn[
  \begin{@twocolumnfalse}
    \maketitle
    \begin{abstract}\noindent
     #1
	~\\\noindent 
	\rule{0.15\textwidth}{0pt}
	\rule{0.05\textwidth}{0.5pt}\rule{0.07\textwidth}{1pt}\rule{0.36\textwidth}{1.5pt}\rule{0.07\textwidth}{1pt}\rule{0.05\textwidth}{0.5pt}
       ~\\\noindent
    \end{abstract}
  \end{@twocolumnfalse}
]
}

\usepackage{soul,xcolor}\setstcolor{red}\definecolor{newc}{rgb}{0.,0.3,0.7}

\definecolor{newerc}{HTML}{009900}

\begin{document} 

\natabstract{
Cosmological observations indicate that 85\% of all matter in the Universe is dark matter (DM), yet its microscopic composition remains a mystery.
One hypothesis is that DM arises from ultralight quantum fields that form macroscopic objects such as topological defects. 
Here we use GPS as a $\sim$\,50,000\,{km} aperture DM detector to search for such defects in the form of domain walls. 
GPS navigation relies on precision timing signals furnished by atomic clocks hosted on board GPS satellites.
As the Earth moves through the galactic DM halo, interactions with topological defects could cause atomic clock glitches that propagate through the GPS satellite constellation at galactic velocities $\sim$\,300\,{km\,s$^{-1}$}. 
Mining 16 years of archival GPS data, we find no evidence for DM in the form of domain walls at our current sensitivity level.
This allows us to improve the limits on certain quadratic scalar couplings  of domain wall DM to standard model particles by several orders of magnitude. 
}

\noindent
Despite the overwhelming cosmological evidence for the existence of dark matter (DM),  there is as of yet no definitive evidence for DM in terrestrial experiments.
Multiple cosmological observations suggest that ordinary matter makes up only about 15\% of the total matter in the universe, with the remaining portion composed of DM~\cite{Bertone2005}.
All the evidence for DM (e.g., galactic rotation curves, gravitational lensing, cosmic microwave background) comes from  galactic or larger scale observations through  the gravitational pull of DM on ordinary matter~\cite{Bertone2005}. 
Extrapolation from the galactic to laboratory scales presents a challenge because of the unknown nature of DM constituents. 
Various theories postulate additional non-gravitational interactions between Standard Model (SM) particles and DM.  
Ambitious programs in particle physics have mostly focused on (so far unsuccessful) searches for WIMP  (Weakly Interacting Massive Particle) DM candidates  with $10-10^3 \, \mathrm{GeV}\,c^{-2}$ masses ($c$ is the speed of light) through their energy deposition in particle detectors~\cite{Liu2017}. 
The null results of the WIMP searches have partially motivated an increased interest in alternative DM candidates, such as ultralight fields.
These fields, in contrast to particle candidates, act as coherent entities on the scale of an individual detector.  

Here we focus on ultralight fields that may cause apparent variations of fundamental constants of nature. 
Such variations in turn lead to shifts in atomic energy levels, which may be measurable by monitoring atomic frequencies~\cite{RosenSci2008,HunLipTam14,GodNisJon14}. 
Such monitoring is performed naturally in atomic clocks, which tell time by locking the frequency of externally generated  electromagnetic radiation to atomic frequencies. 
Here, we analyze time as measured by atomic clocks on board Global Positioning System (GPS) satellites to search for DM-induced transient variations of fundamental constants~\cite{DereviankoDM2014}. 
In effect  we use the GPS constellation  as a $\sim$\,50,000\,km-aperture DM detector.
Our DM search is one example of using GPS for fundamental physics research. 
Another recent example includes placing limits on gravitational waves~\cite{Aoyama2014}.

GPS works by broadcasting microwave signals from nominally 32 satellites in medium-Earth orbit. 
The signals are driven by an atomic clock (either based on Rb or Cs atoms) on board each satellite.
By measuring the carrier phases of these signals with a  global network of specialised GPS receivers, the geodetic community can position stations at the 1\,mm level for purposes of investigating plate tectonics and geodynamics~\cite{Blewitt2015307}.  As part of this data processing,  the time differences between satellite and station clocks are determined with $<0.1\, \mathrm{ns}$ accuracy~\cite{RaySenior2005}.
Such high-quality timing data for at least the past decade are publicly available and are routinely updated.
Here we analyze  data from the Jet Propulsion Laboratory~\cite{MurphyJPL2015}.
A more detailed overview of the GPS architecture and data processing relevant to our search is given in section~S.1 of the Supplementary Material.

\begin{figure}[t]
\centering
\includegraphics[height=0.33\textwidth]{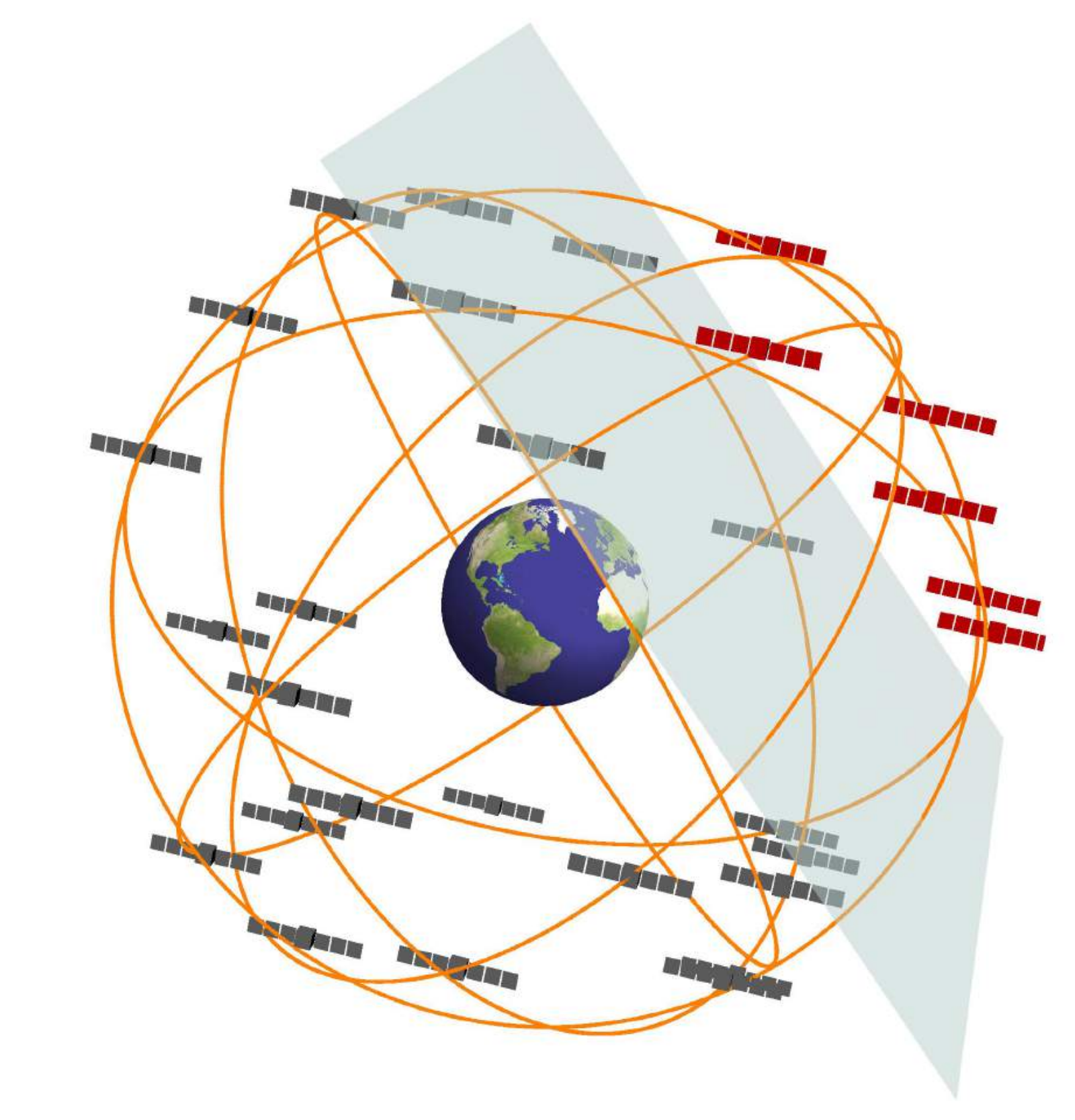}
	\caption{ \small 
	As a domain wall sweeps through the GPS constellation at galactic velocities, $v_g\sim300\,${km\,s$^{-1}$},  
it perturbs  the atomic clocks on board the satellites causing a correlated propagation of glitches through the network.
}
\label{fig:walla}
\end{figure}

The large aperture of the GPS network is well suited to search for macroscopic DM objects or ``clumps''. Examples of clumpy DM are numerous: topological defects (TDs)~\cite{Kibble1980,Vilenkin1985}, $Q$-balls~\cite{Coleman1985,Kusenko2001,Lee1989}, solitons~\cite{Marsh2015,Schive2014}, axion stars~\cite{Hogan1988,Kolb1993}, and other stable objects formed due to dissipative interactions in the DM sector. 
For concreteness, we interpret our results in terms of TDs.
Each TD type (monopoles, strings, or domain walls) would exhibit a transient in GPS data with a distinct  signature. General signature matching for the vast set of GPS data has proven to be computationally expensive and is in progress.
Here, we report the results of the search for domain walls, quasi-2D cosmic structures, since domain walls would leave the simplest DM signature in the data. An example of a domain wall crossing is shown in Fig.~\ref{fig:walla}. 
While we interpret our results in terms of domain wall DM, we remark that our search applies equally to the situation where walls are closed on themselves, forming a bubble that has transverse size significantly exceeding the terrestrial scale. 
The galactic structure formation in that case may occur as per conventional cold dark matter theory\cite{Blumenthal1984}, since from the large distance perspective the bubbles of domain walls behave as point-like objects.

Topological defects may be formed during the cooling of the early universe through a spontaneous symmetry breaking phase transition~\cite{Kibble1980,Vilenkin1985}. 
Technically, this requires the existence of hypothesised self-interacting DM  fields,  $\varphi$. 
While the exact nature of TDs is model-dependent, the spatial scale  of the DM object, $d$, is generically given by the Compton wavelength of the particles that make up the DM field $d = \hbar/(m_\varphi c)$, where $m_\varphi$ is the field particle mass, and $\hbar$ is the reduced Plank constant.
The fields that are of interest here are ultralight: for an Earth-sized object the mass scale is $m_\varphi\sim10^{-14}\,\mathrm{eV}\,c^{-2}$, hence the probed parameter space is complementary to that of WIMP searches~\cite{Liu2017}, as well as searches for other DM candidates~\cite{Bozek2015,Arvanitaki2011,Arvanitaki2014b}.
Searches for TDs have been performed via their gravitational effects, including gravitational lensing~\cite{Schneider1992,Vilenkin1994,Cline2001}.
Limits on TDs have been placed by Planck~\cite{Planck2014-TD} and BICEP2~\cite{BICEP2-2014} from fluctuations in the cosmic microwave background. 
So far the existence of TDs is neither confirmed nor ruled out.
The past few years have brought several proposals for TD  searches via their {\em non}-gravitational signatures~\cite{DereviankoDM2014,Pospelov2013,Pustelny2013,Hall2016,StadnikDefects2014,Kalaydzhyan2017}.

\begin{figure}[t]
\centering
\includegraphics[height=0.33\textwidth]{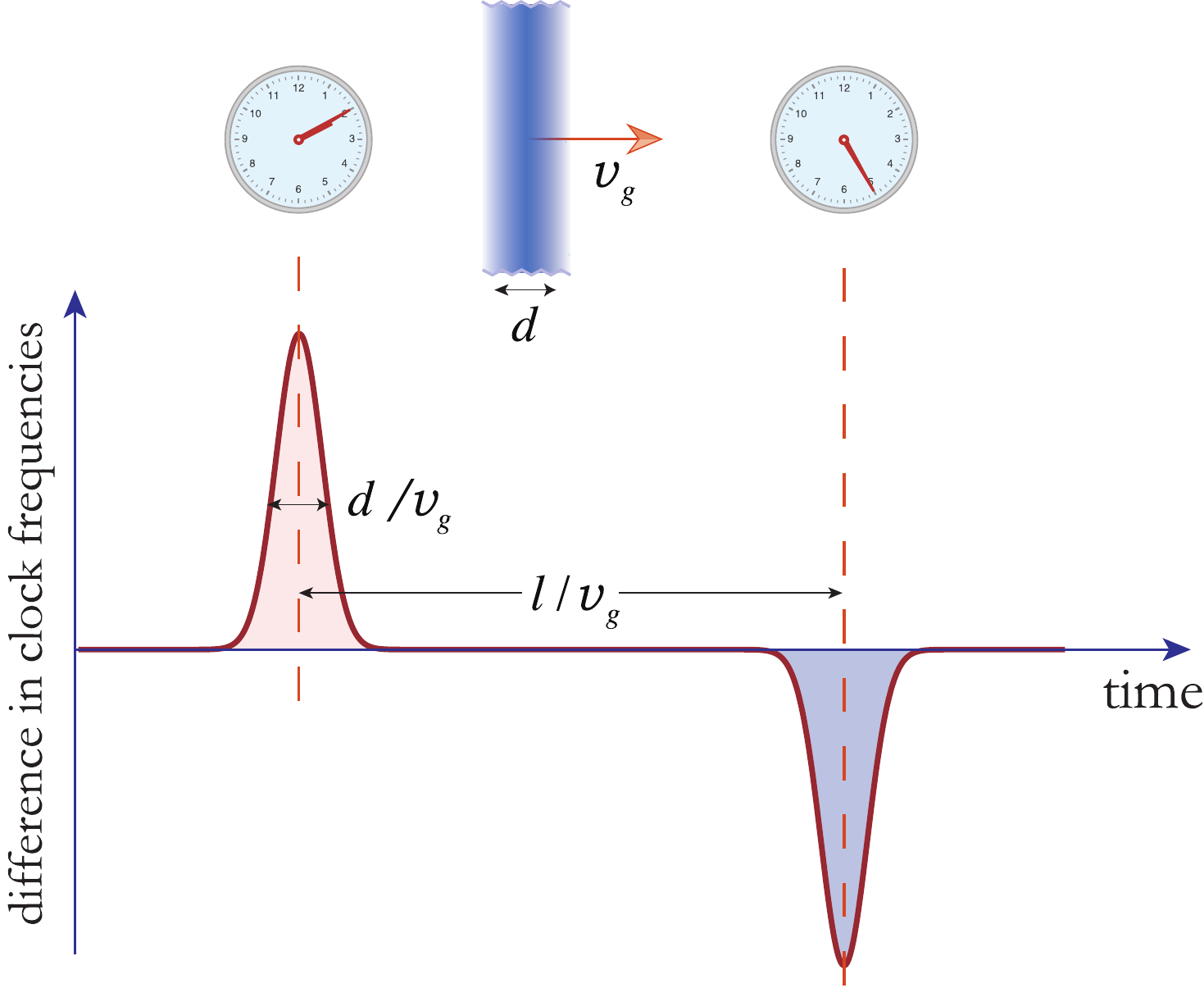}
	\caption{ \small 
	The time-dependence of the DM-induced frequency difference  between two identical clocks separated by distance $l$. 
         The time delay in the signals encodes the kinematics of the DM object.
}
\label{fig:wallb}
\end{figure}


\begin{figure*}
\centering
\includegraphics[height=0.275\textwidth]{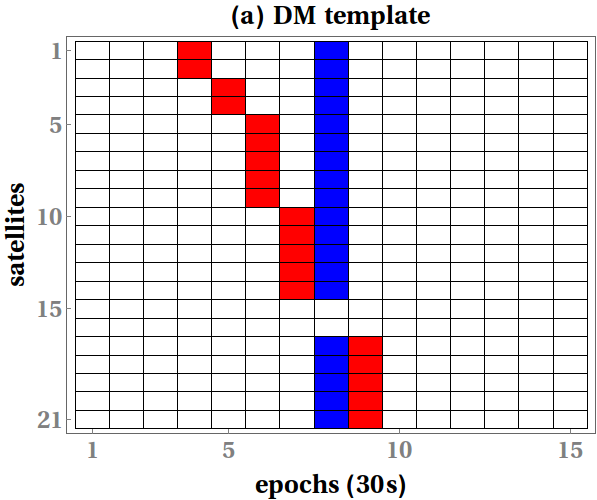}~~~~
\includegraphics[height=0.275\textwidth]{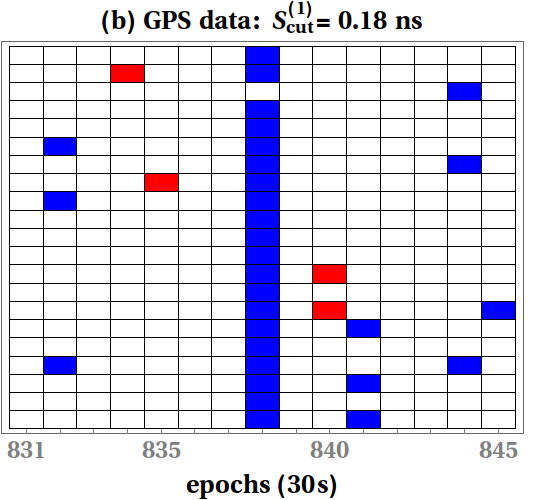}
\includegraphics[height=0.275\textwidth]{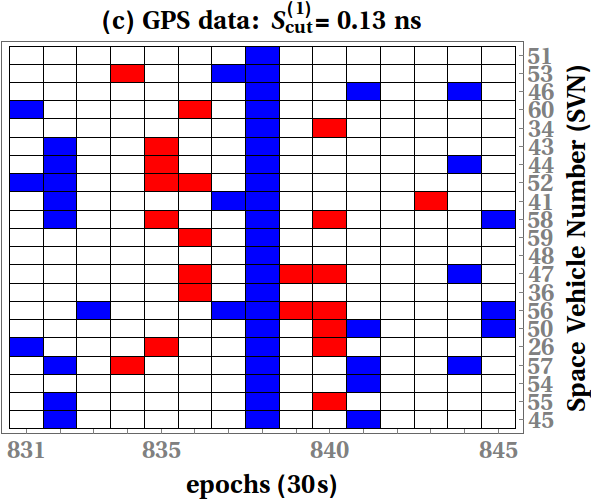}
	\caption{ \small 
(a) One of the expected  pseudo-frequency $S^{(1)}$  signatures for a thin domain wall. Red (blue) tiles indicate positive (negative) DM-induced frequency excursions, while white tiles mark the absence of the signal (c.f.~Fig.~\ref{fig:wallb}).
In this example, the satellites are listed in the order they were swept (though in general the order depends on the incident direction of the DM object and is not known {\em a priori}), and $\Gamma_{\rm eff}>0$ in Eq.~(\ref{Eq:X-shift}). 
The slope of the red line encodes the incident velocity of the wall.
The reference clock was swept within the 30\,s leading to epoch 8.
Satellites 15 and 16 do not record any frequency excursions, since they are spatially close the reference clock and are swept within the same 30\,s period.
Panels (b) and (c) show $S^{(1)}$ GPS satellite data streams for all operational Rb clocks for 21 May, 2010 for a 15 epoch window. 
Red tiles show data points with $S^{(1)}>S^{(1)}_\mathrm{cut} $, and the blue depict $S^{(1)}<-S^{(1)}_\mathrm{cut}$, 
with $S^{(1)}_\mathrm{cut} = 0.18 \, \mathrm{ns}$ and $0.13 \, {\rm ns}$, respectively.
At the $0.13\,{\rm ns}$ level, panel (c), this data window would be flagged as a potential event, but not at the $0.18\,{\rm ns}$ level shown in panel (b).
In this case, the potential event (c) is excluded because the reference clock experiences a much larger perturbation than the rest of the clock network.
}
\label{fig:tiles}
\end{figure*}

We employ the known properties of the DM halo to model the statistics of encounters of the Earth with TDs.
Direct measurements~\cite{Bovy:2012tw} of the local dark matter density give $0.3\pm0.1 \, \mathrm{GeV}\,\mathrm{cm}^{-3}$,
and we adopt the value of $\rho_\mathrm{DM} \approx 0.4 \, \mathrm{GeV}\,\mathrm{cm}^{-3}$ for definitiveness. 
According to the standard halo model, in the galactic rest frame the velocity distribution of DM objects is isotropic and quasi-Maxwellian, with dispersion~\cite{Freese2013} $v\simeq 290\,${km\,s$^{-1}$} and a cut-off above the galactic escape velocity of $v_{\rm esc} \simeq 550\,${km\,s$^{-1}$}.
The Milky Way rotates through the DM halo with the Sun moving at $\sim$\,$220\,${km\,s$^{-1}$} towards the Cygnus constellation. 
For the goals of this work we can neglect the much smaller orbital velocities of the Earth around the Sun ($\sim$\,$30\,{\rm km\,s}^{-1}$) and GPS satellites around the Earth ($\sim$\,$4\,{\rm km\,s}^{-1}$).
Thereby one may think of a TD ``wind''  impinging upon the Earth, with typical relative  velocities $v_g \sim 300 \,${km\,s$^{-1}$}.
Assuming the standard halo model, the vast majority of events ($\sim$\,95\%) would come from the forward-facing hemisphere centred about the direction of the Earth's motion through the galaxy, with typical transit times through the GPS constellation of about three minutes.
A positive DM signal can be visualised as a coordinated propagation of clock ``glitches'' at galactic velocities through the GPS constellation, see Fig.~\ref{fig:wallb}. 
Note that we make an additional assumption that the distribution of wall velocities is similar to the standard halo model, which is expected if the gravitational force is the main force governing wall dynamics within the galaxy. 
However, even if this distribution is somewhat different, the qualitative feature of a TD ``wind'' is not expected to change.
The powerful advantage of working with the network is that non-DM clock perturbations do not mimic this signature. 
The only systematic effect that has propagation velocities comparable to $v_g$ is  the solar wind~\cite{SolarWindBook}, an effect that is simple to exclude based on the distinct directionality from the Sun and the fact that the solar wind does not affect the satellites in the Earth's shadow.

As the nature of non-gravitational interactions of DM with ordinary matter is unknown, we take a phenomenological approach that respects the Lorentz and local gauge invariances. 
We consider quadratic scalar interactions between the DM objects and clock atoms that can be parameterised in terms of shifts in the effective values of fundamental constants~\cite{DereviankoDM2014}.  
The relevant combinations of fundamental constants include
 $\alpha = e^2/\hbar c \approx1/137$, the dimensionless electromagnetic fine-structure constant ($e$ is the elementary charge), the ratio $m_q/\Lambda_{\rm QCD}$ of the light quark mass to the quantum chromodynamics (QCD) energy-scale, and $m_e$ and $m_p$, the electron and proton masses.
With the quadratic scalar coupling, the relative change in the local value for each such fundamental constant is  proportional to the square of the DM field 
\begin{equation}
\frac{\delta X (\mathbf{r},t)}{X} =  \Gamma_X  \, \varphi(\mathbf{r},t)^2,
\label{Eq:X-shift}
\end{equation} 
where 
$\Gamma_X $ is the coupling  constant between dark and ordinary matter, with $X=\alpha, \, m_e,\, m_p, \, m_q/\Lambda_{\rm QCD}$ (see section~S.2 of the Supplementary Information for details).

As the DM field vanishes outside the TD, the apparent variations in the fundamental constants occur only when the TD overlaps with the clock.
This temporary shift in the fundamental constants leads in-turn to a transient shift in the atomic  frequency referenced by the clocks, which may be measurable by monitoring atomic frequencies~\cite{RosenSci2008,HunLipTam14,GodNisJon14}. 
The frequency shift can be expressed as
\begin{equation}
\frac{\delta \omega (\mathbf{r},t)}{\omega_c} = \sum_{X} K_X\frac{\delta X (\mathbf{r},t)}{X},
 \label{Eq:ClockFreqPert}
\end{equation} 
where $\omega_c$ is the unperturbed clock frequency and $K_X$ are known  coefficients of sensitivity to effective changes in the constant $X$  for a particular clock transition~\cite{FlambaumCJP2009}.
It is worth noting, that the values of the sensitivity coefficients $K_X$ depend on  experimental realisation. 
Here we compare spatially separated clocks (to be contrasted with the conventional frequency ratio comparisons~\cite{RosenSci2008,HunLipTam14,GodNisJon14}), and thus our used values of $K_X$ somewhat differ from Ref.~\citenum{FlambaumCJP2009}; full details are presented in section~S.2 of the Supplementary Information.
For example, for the microwave frequency $^{87}$Rb clocks on board the GPS satellites,
the sensitivity coefficients are
\begin{equation}
\label{eq:KRb}
\frac{\delta \omega}{\omega_c}({\rm Rb})
=({4.34}\,{\Gamma_\alpha}-{0.019}\,{\Gamma_q}+\Gamma_{e/p})\,\varphi^2 \equiv \Gamma_{\rm eff}^{\rm (Rb)}\,\varphi^2,
\end{equation}
where we have introduced the short-hand notation $\Gamma_q\equiv\Gamma_{m_q/\Lambda_{\rm QCD}}$ and $\Gamma_{e/p}\equiv2\Gamma_{m_e} - \Gamma_{m_p}$,
and the effective coupling constant 
$\Gamma_\mathrm{eff} \equiv \sum_X K_X \Gamma_X $.

From Eqs.~(\ref{Eq:X-shift}) and (\ref{Eq:ClockFreqPert}), the  extreme TD-induced frequency excursion, 
$\delta {\omega}_\mathrm{ext}$,  
is related to the field amplitude $\varphi_{\rm max}$  inside the defect as
$\delta {\omega}_\mathrm{ext} =   \Gamma_\mathrm{eff} {\omega _c}{\varphi_{\rm max}^2}$. 
Further, assuming that a particular TD type saturates the DM energy density, we have~\cite{DereviankoDM2014} 
$\varphi_{\rm max}^{2}=\hbar c \rho_\mathrm{DM} \mathcal{T} v_{g} d$.
Here, $\mathcal{T}$ is the average time between consecutive encounters of the clock with DM objects, which, for a given $\rho_\mathrm{DM}$, depends on the energy density inside the defect~\cite{DereviankoDM2014}
$
\rho_{\rm inside} = \rho_{\rm DM} {\mathcal{T} v_g}/{d}.
$
Thus the expected DM-induced fractional frequency excursion reads 
\begin{equation}
\frac{{\delta \omega_\mathrm{ext} }} {\omega _c}  = \Gamma_\mathrm{eff}  \,  { \hbar c \, \rho _\mathrm{DM}}  {v_g}\, \mathcal{T}  d ,
 \label{Eq:Limit}
\end{equation}
which is valid for TDs of any type (monopoles, walls, and strings).
The frequency excursion is positive for $\Gamma_\mathrm{eff}>0$, and negative for $\Gamma_\mathrm{eff}<0$.

The key qualifier for the preceding equation~(\ref{Eq:Limit}) is that one must be able to distinguish between the clock noise and DM-induced frequency excursions. Discriminating between the two sources  relies on measuring time delays between DM events at network nodes. 
Indeed, if we
consider a pair of spatially separated clocks 
 (Fig.~\ref{fig:wallb}),
the DM-induced frequency shift~(\ref{Eq:ClockFreqPert}) translates into a distinct pattern. 
The velocity of the sweep is encoded in the time delay between two DM-induced spikes and it must lie within the boundaries predicted by the standard halo model.
Generalization to the multi-node network is apparent (see GPS-specific discussion below). The distributed response of the network  encodes the spatial structure and kinematics of the DM object, and its coupling to atomic clocks.

Working with GPS data introduces several peculiarities into the above discussion (see section~S.1 of the Supplementary Information for details). 
The most relevant is that the available GPS clock data are clock  biases (i.e.,  time differences between the satellite and reference clocks)  $S^{(0)}(t_k)$  sampled at times (epochs) $t_k$ every 30 seconds.
Thus  we cannot access the continuously sampled clock frequencies as in 
Fig.~\ref{fig:wallb}. 
Instead, we formed discretised pseudo-frequencies 
$S^{(1)}(t_k) \equiv S^{(0)}(t_{k}) -  S^{(0)}(t_{k-1})$. 
Then the  signal is especially simple if the DM object transit time through a given clock, $d/v_g$, is smaller than the 30-second epoch interval 
(i.e., ``thin'' DM objects with $d \lesssim 10^4 \, \mathrm{km}$, roughly the size of the Earth),
since in this case  $S^{(1)}$ collapses into a solitary spike at $t_k$ if the DM object was encountered during the 
$(t_{k-1}, t_{k})$ interval. 
The exact time of interaction within this interval is treated as a free parameter.

\begin{figure}[t]
\centering
	\includegraphics[width=0.4\textwidth]{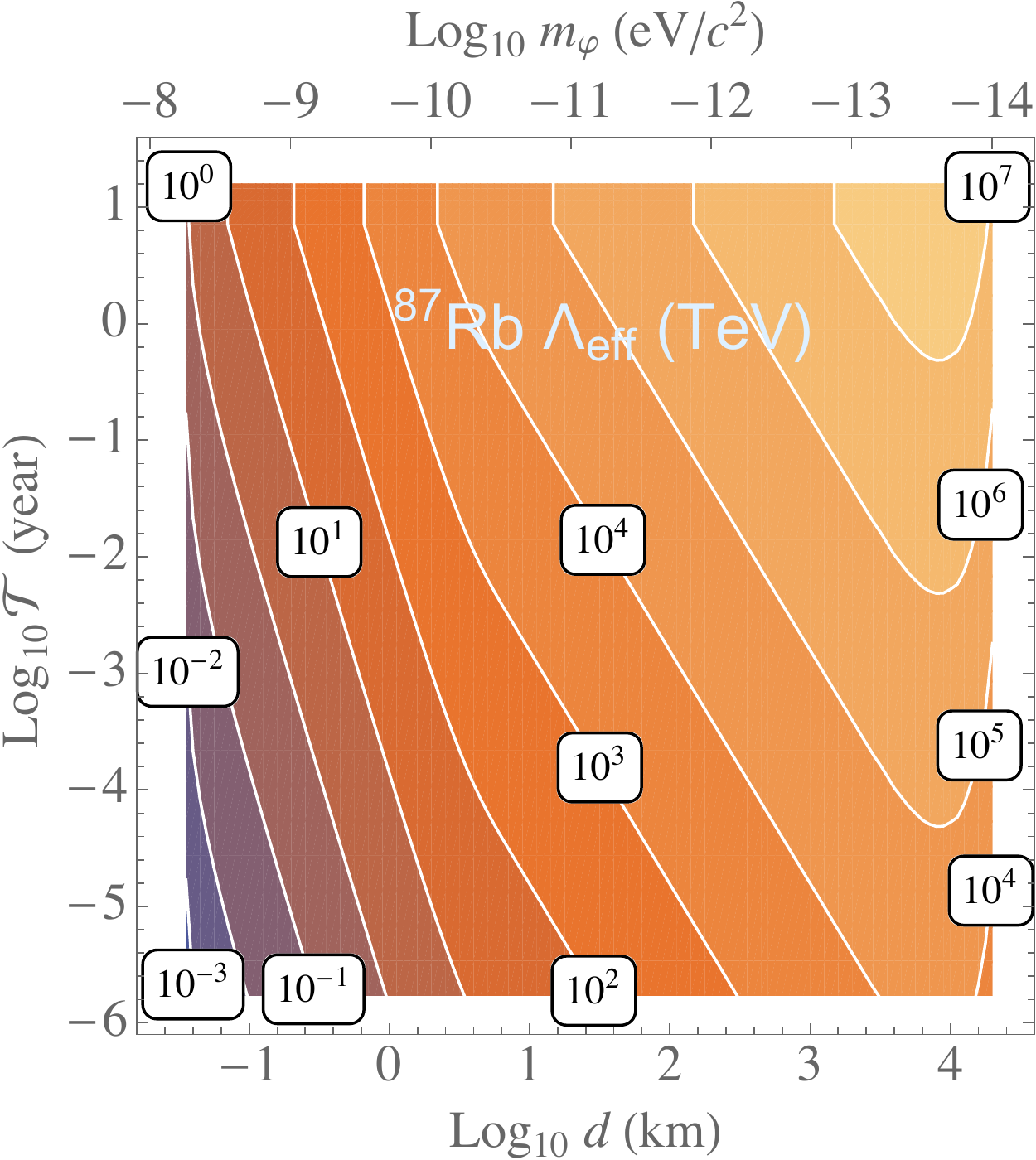}
	\caption{\small 
Contour plot showing the 90\% C.L. exclusion limits on the effective energy scale  $\Lambda_{\rm eff}$ from the GPS Rb sub-network as a function of the wall width, $d$, and average time between encounters with domain walls, $\mathcal{T}$.
	} 
\label{fig:limits-eff}
\end{figure}

One of the expected $S^{(1)}$ signatures for a thin domain wall propagating through the GPS constellation is shown in Fig.~\ref{fig:tiles}(a). 
This signature was generated  for a domain wall  incident with $v= 300\,${km\,s$^{-1}$}  from the most probable direction.
The derivation of the specific expected domain wall signal is presented in section~S.3 of the Supplementary Information.
Since the DM response of Rb and Cs satellite clocks can differ due to their distinct effective coupling constants $\Gamma_\mathrm{eff}$, we treated the Cs and Rb satellites as two  sub-networks, and performed the analysis separately.
Within each sub-network we chose the clock on board the most recently launched satellite as the reference because, as a rule, such clocks are the least noisy among all the clocks in orbit.

To search for domain wall signals, we analyzed the $S^{(1)}$ GPS data streams in two stages.
At the first stage, we scanned all the data from May 2000 to October 2016 searching for the most general patterns associated with a domain wall crossing, without taking into account the order in which the satellites were swept.
We required at least 60\% of the clocks to experience a frequency excursion at the same epoch, which would correspond to when the wall crossed the reference clock (vertical blue line in Fig.~\ref{fig:tiles}(a)).
This 60\% requirement is a conservative choice based on the GPS constellation geometry, and ensures sensitivity to walls with relative speeds of up to $v\lesssim700\,{\rm km\,s}^{-1}$.
Then, we checked if these clocks also exhibit a frequency excursion of similar magnitude (accounting for clock noise) and  opposite sign anywhere else within a given time window (red tiles in Fig.~\ref{fig:tiles}(a)).
Any epoch for which these criteria were met was counted as a ``potential event''.
We considered time windows corresponding to sweep durations through the GPS constellation of up to 15,000\,{\rm s}, which is sufficiently long to ensure sensitivity to walls moving at relative velocities $v\gtrsim 4\,{\rm km\, s}^{-1}$  
(given that $<$\,0.1\% of DM objects are expected to move with velocities outside of this range).
The full details of the employed search technique are presented in section~S.4 of the Supplementary Information.

The tiled representation of the GPS data stream depends on the chosen signal cut-off $S^{(1)}_\mathrm{cut}$ (see Fig.~\ref{fig:tiles}). 
We systematically decreased the cut-off values and repeated the above procedure.
Above a certain threshold, $S^{(1)}_\mathrm{thresh}$, no potential events were seen.
This process is demonstrated for a single arbitrarily chosen data window in Figs.~\ref{fig:tiles}(b) and \ref{fig:tiles}(c).
The thresholds for the Rb and Cs subnetworks above which no potential events were seen are
$S^{(1)}_\mathrm{thresh} (\mathrm{Rb}) = 0.48\,{\rm ns}$ and $S^{(1)}_\mathrm{thresh} (\mathrm{Cs}) =  0.56\,{\rm ns}$ for $v \approx 300\,${km\,s$^{-1}$} sweeps.

The second stage of the search involved analyzing the ``potential events'' in more detail, so that we may elevate their status to ``candidate events'' if warranted by the evidence.
We examined a few hundred potential events that had $S^{(1)}$ magnitudes just below $S^{(1)}_\mathrm{thresh}$, by matching the data streams against the expected patterns; one such example is shown in Fig.~\ref{fig:tiles}(a).
At this second stage, we accounted for the ordering and time at which each satellite clock was affected.
The velocity vector and wall orientation were treated as free parameters within the bounds of the standard halo model.
As a result of this pattern matching, we found that none of these events were consistent with domain wall DM, thus we have found no candidate events at our current sensitivity.
Analysing numerous potential events well below $S^{(1)}_{\rm thresh}$ has proven to be substantially more computationally demanding, and is beyond the scope of the current work.

\begin{figure*}
\centering
	\includegraphics[width=0.49\textwidth]{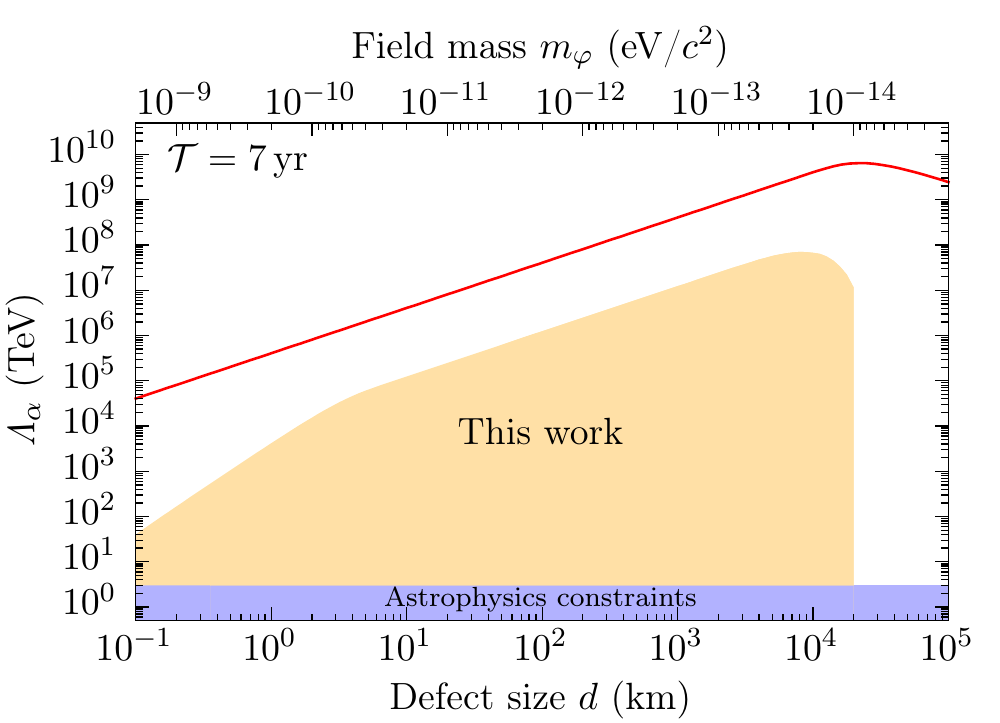}
	\includegraphics[width=0.49\textwidth]{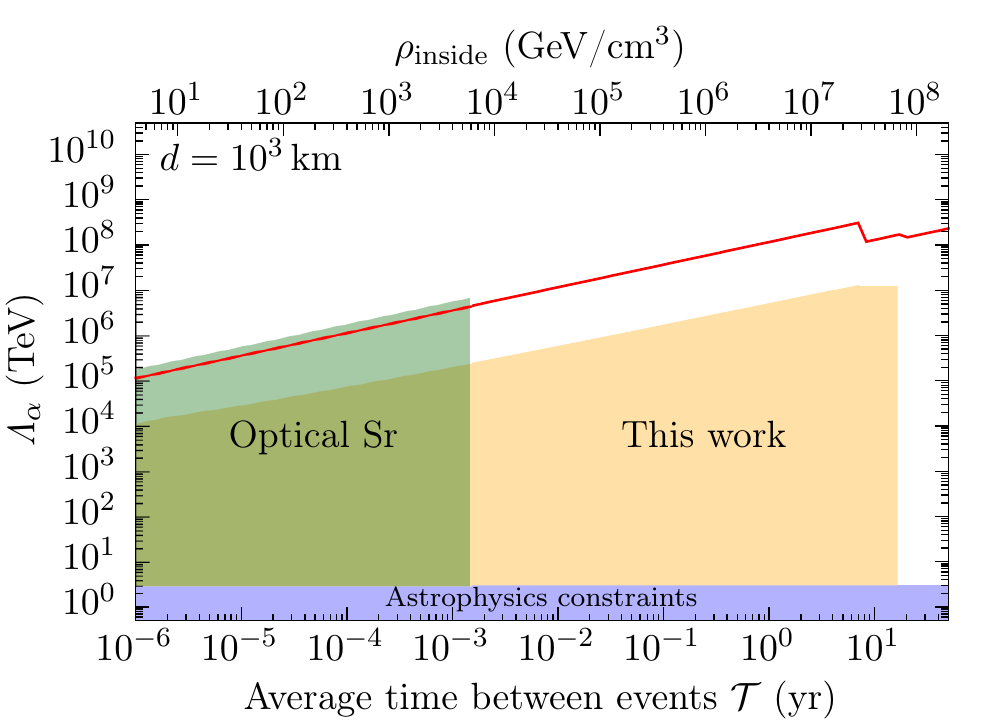}
	\caption{\small  Limits (90\% C.L.) on the energy scale $\Lambda_\alpha$, as a function of the wall width, $d$ (left panel), and average time between encounters, $\mathcal{T}$ (right panel). 
The shaded yellow region shows the GPS limits from this work (assuming ${\Gamma_{\alpha}}\gg{\Gamma_{q,e/p}}$), the shaded green region shows the limits derived from an optical Sr clock~\cite{Wcislo2016}, and the shaded blue region shows the astrophysical bounds~\cite{Olive2008}.  
The solid red line shows the potential discovery reach using the global network of GPS microwave atomic clocks.
For $\mathcal{T}\lesssim7\,{\rm yr}$, the GPS reach is limited by the modern Rb block IIF satellite clocks~\cite{Griggs2015} ($\sigma_y(30\,{\rm s})\sim 10^{-11}$), and for $\mathcal{T}\gtrsim7\,{\rm yr}$, the reach is limited by the older Rb (block IIR, IIA and II) GPS clocks ($\sigma_y(30\,{\rm s})\sim 10^{-10}$).
Compared to more accurate optical clocks, microwave clocks provide additional  sensitivity to  $\Lambda_q$ and $\Lambda_{e/p}$(optical clocks only have sensitivity to $\Lambda_\alpha$).
}
\label{fig:limits-alpha}
\end{figure*}

Since we did not find evidence for encounters with domain walls at our current sensitivity, there are two possibilities: (i) DM of this nature does not exist, or (ii) the DM signals are below our sensitivity. In the latter case we may constrain the possible range of the coupling strengths $\Gamma_\mathrm{eff}$. 
For the discrete pseudo-frequencies, and considering the case of thin domain walls, Eq.~(\ref{Eq:Limit}) becomes  
\begin{equation}
\left| \Gamma_\mathrm{eff} \right|  <  \frac{S^{(1)}_{\rm thresh} } {\hbar c \sqrt{\pi}\rho_{\rm DM} \, \mathcal{T} d^2} \,.
\label{Eq:GammaEffLimits}
\end{equation}
Our technique is not equally sensitive to all values for the wall widths, $d$, or average times between collisions, $\mathcal{T}$. 
This is directly taken into account by introducing a sensitivity function that is combined with Eq.~(\ref{Eq:GammaEffLimits}) to determine the final limits at the 90\% confidence level.
For example, the smallest width is determined by the servo-loop time of the GPS clocks, i.e.~by how quickly the clock responds to the changes in atomic frequencies. 
In addition, we are sensitive to events that occur less frequently than once every $\sim$\,150\,s (so the expected patterns do not overlap), which places the lower bound on $\mathcal{T}$.
Further, we incorporate the expected event statistics into Eq.~(\ref{Eq:GammaEffLimits}). 
Details can be found in section~S.5 of the Supplementary Information.
 
 Our results are presented in Fig.~\ref{fig:limits-eff}. 
To be consistent with previous literature~\cite{DereviankoDM2014,Wcislo2016}, the  limits are presented  for  the effective energy scale $\Lambda_\mathrm{eff} \equiv 1/\sqrt{ \left| \Gamma_\mathrm{eff} \right|}$.
Further, on the assumption that the coupling strength $\Gamma_\alpha$ dominates over the other couplings in the linear combination (\ref{eq:KRb}), we  place limits on $\Lambda_\alpha$.
The resulting limits are  shown in 
Fig.~\ref{fig:limits-alpha}, together with existing constraints~\cite{Wcislo2016,Olive2008}.
For certain parameters, our limits exceed the $10^7\,{\rm TeV}$ level; 
 astrophysical limits~\cite{Olive2008} on $\Lambda_\alpha$, which come from stellar and supernova energy-loss observations \cite{Raffelt1999,Hirata1988}, have not exceeded $\sim$\,$10\,{\rm TeV}$. 

The derived constraints on $\Lambda_\alpha$ can be translated into a limit on the transient variation of the fine-structure constant,
\begin{equation}
	\frac{\delta\alpha}{\alpha}=\hbar c \rho_{\rm DM} v_g \frac{\mathcal{T}\,d }{\Lambda^2_\alpha K_\alpha},
\end{equation}
which for $d=10^4\,{\rm km}$ corresponds to
${\delta\alpha}/{\alpha}\lesssim 10^{-12}$.
Due to the scaling of the constraints on $\Lambda_X$, this result is independent of $\mathcal{T}$, and scales inversely with $d$ (within the region of applicability).
It is worth contrasting this constraint with results from the searches for slow linear drifts of fundamental constants. 
For example, the search~\cite{GodNisJon14}  resulting in the most stringent limits on long-term drifts of $\alpha$ was carried out over a year and led to 
$\frac{\delta\alpha}{\alpha} \lesssim 3 \times 10^{-17}$.
Such limits may apply only for very thick walls $d \gg v_g \times 1\,{\rm yr} \sim 10^{10} \,{\rm km}$, which are outside our present discovery reach.

\begin{figure*}
\centering
	\includegraphics[width=0.49\textwidth]{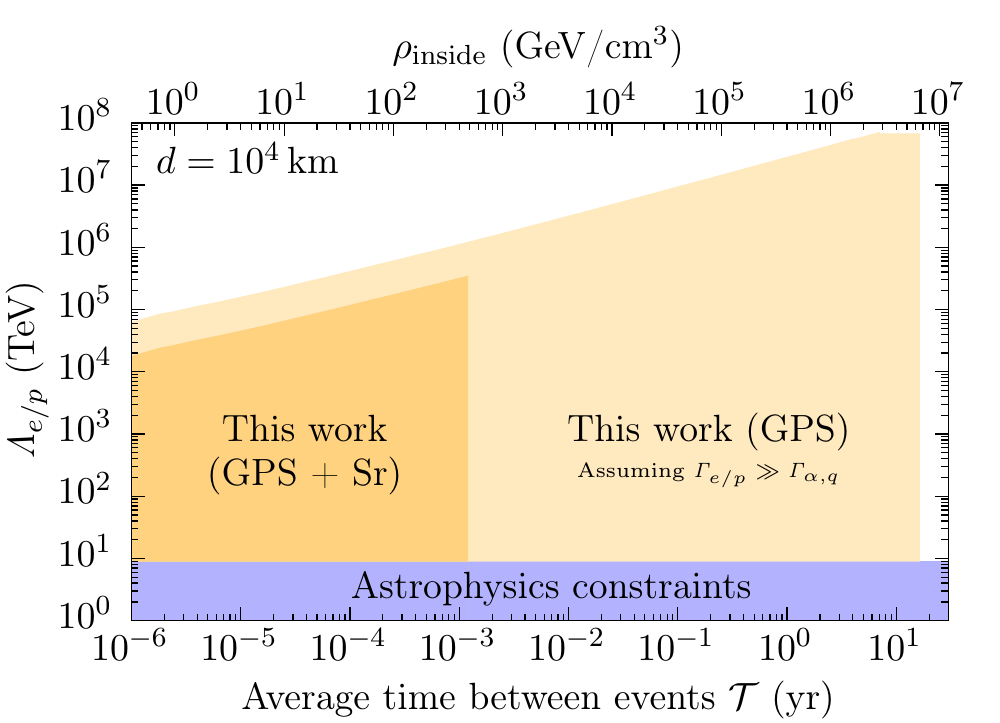}
	\includegraphics[width=0.49\textwidth]{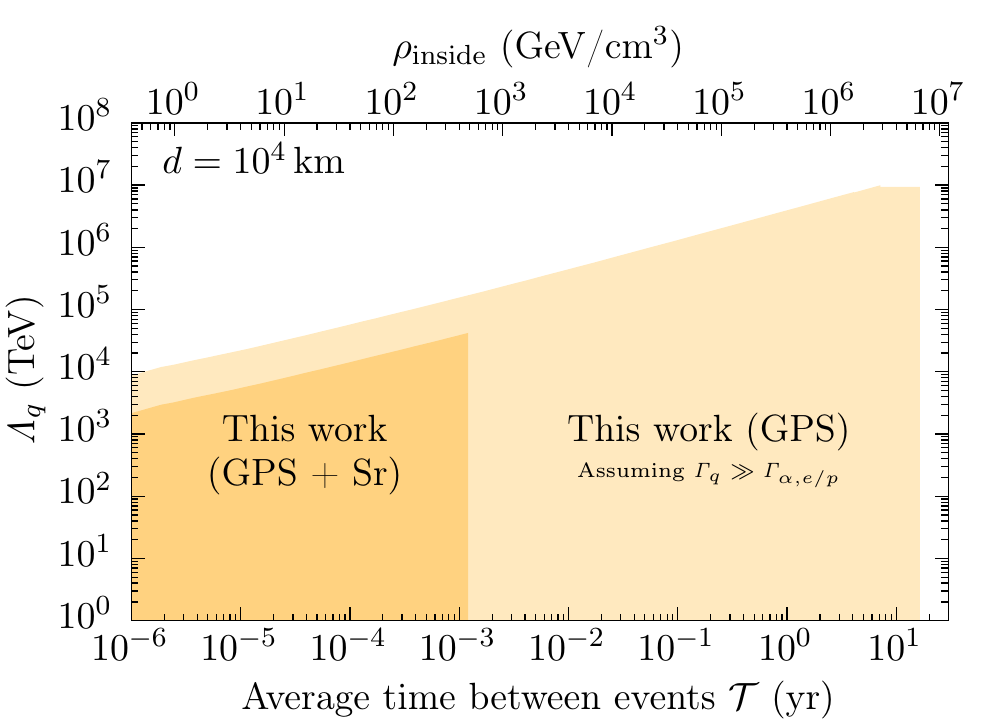}
	\caption{\small  Limits (90\% C.L.) on the energy scale $\Lambda_{e/p}$ (left panel) and $\Lambda_q$ (right panel), as a function of the average time between encounters, $\mathcal{T}$, for constant $d=10^4\,{\rm km}$. 
The lighter yellow regions are the limits from the Rb GPS sub-network in the assumption that the respective couplings dominate the interactions. The darker region combines our GPS limits (from the Rb and Cs sub-networks) with the limits on $\Lambda_\alpha$ from the Sr optical clock constraints~\cite{Wcislo2016} to place assumption-free limits on $\Lambda_{e/p}$  and $\Lambda_{q}$.
The blue region shows the astrophysical bounds for $\Lambda_{e/p}$; note that $\Lambda_{q}$ was previously unconstrained~\cite{Olive2008}. 
	} 
\label{fig:limits-epq}
\end{figure*}

Further, by combining our results from the Rb and Cs GPS sub-networks with the recent limits on $\Lambda_\alpha$ from an optical Sr clock~\cite{Wcislo2016}, we  also place independent limits on 
$\Lambda_{e/p}$, and $\Lambda_{q}$;
for details, see section~S.6 of the Supplementary Information.
These limits are presented in Fig.~\ref{fig:limits-epq} as a function of the average time between events.
For certain values of the $d$ and $\mathcal{T}$ parameters, we improve current bounds on $\Lambda_{e/p}$ by a factor of $\sim 10^5$ and for the first time establishes limits on $\Lambda_{q}$.


While we have improved the current constraints on DM-induced transient variation of fundamental constants by several orders of magnitude, it is possible that DM events remain undiscovered in the data noise.  
Our current threshold  $S^{(1)}_{\rm thresh}$ is larger than the GPS data noise  by a factor of $\sim 5-20$, depending on which clocks/time periods are examined. 
By applying a more sophisticated statistical approach with greater computing power, we expect to improve our sensitivity by up to two orders of magnitude.
Indeed, the sensitivity of the search is statistically determined by the number of clocks in the network $N_{\rm clocks}$ and the Allan deviation~\cite{Barnes1971}, $\sigma_y(\tau_0)$, evaluated at the data sampling interval $\tau_0=30\,{\rm s}$ reads,
\begin{equation}\label{eq:reach}
S^{(1)}  \gtrsim \frac{\sigma_y(\tau_0)\,\tau_0}{\sqrt{N_{\rm clocks}}},
\end{equation}
or, combining with Eq.~(\ref{Eq:GammaEffLimits}),
\begin{equation}\label{eq:limitform}
\Lambda_X \lesssim d\sqrt{ \frac{\hbar c \rho_{\rm DM} \mathcal{T}\,K_X\,\sqrt{N_{\rm clocks}}}{\sigma_y(\tau_0)\,\tau_0} }.
\end{equation}
Note that this estimate differs from that in Ref.~\citenum{DereviankoDM2014}, since while ariving at Eq.~\ref{eq:reach}, we assumed a more realistic white frequency noise (instead of white phase noise).
The projected discovery reach of GPS data analysis is presented in Fig.~\ref{fig:limits-alpha}.

Prospects for the future include incorporating substantially improved clocks on next-generation satellites, increasing the network density with other Global Navigation Satellite Systems, such as European Galileo, Russian GLONASS, and Chinese BeiDou, and including networks of laboratory clocks~\cite{Wcislo2016,Riehle2017}.
Such an expansion can bring the total number of clocks to $\sim$\,100.
Moreover, the GPS search can be extended to other TD types (monopoles and strings), as well as different DM models, such as virialized DM fields~\cite{Kalaydzhyan2017,DereviankoVULF2016}.

In summary, by using the GPS as a dark matter detector, we have substantially improved the current limits on DM domain wall induced transient variation of fundamental constants. 
Our approach relies on mining an extensive set of archival data, using existing infrastructure. 
As the direct DM searches are widening to include alternative DM candidates, it is anticipated that the mining of time-stamped archival data, especially from laboratory precision measurements, will play an important role in verifying or excluding predictions of various DM models~\cite{BudkerPT2015}.
In the future, our approach can be used for a DM search with nascent networks of laboratory atomic clocks that are orders of magnitude more accurate than the current GPS clocks~\cite{Riehle2017}.

~\\[-1.1cm]

	~\\ 
	\noindent\rule{0.01\textwidth}{0pt}
	\rule{0.05\textwidth}{0.5pt}\rule{0.35\textwidth}{1.0pt}\rule{0.05\textwidth}{0.5pt}
	\\[-1.75cm]

{
\small

}
~\\[-0.5cm]
{\small~\\
{\bf Acknowledgements}
We thank A.~Sushkov and Y.~Stadnik for discussions, and J.~Weinstein for his comments on the manuscript.
We acknowledge the International GNSS Service for the GPS data acquisition.
We used JPL's GIPSY software for the orbit reference frame conversion.
This work was supported in part by the US National Science Foundation grant PHY-1506424.
}

~\\
\noindent
{\em Data Availability---} We used publicly available GPS timing and orbit data for  the past 16 years from the Jet Propulsion Laboratory~\cite{MurphyJPL2015}. 

~\\
\noindent
{
\small{$^\ast$andrei@unr.edu.}
}

\setcounter{enumiv}{30}

\natabstract{
{\centering
\section*{%
{\em Supplementary Information}~\\~\\}
}
\noindent{Here we present details of the GPS architecture, data acquisition, and data analysis.
Section~\ref{sec:gpsdata} presents an overview of the relevant aspects of GPS.
Section~\ref{sec:GPSTDSM} outlines the relevant theoretical background and presents the details of the atomic clock response to passing topological defect dark matter.
Section~\ref{sec:DomainWall} discusses the sought domain wall dark matter signals, and section~\ref{sec:window} presents the details of our search method for their transient signatures in the GPS atomic clock data.
Section~\ref{sec:sensitivity} describes the sensitivity of our approach to different regions of the parameter space, and section~\ref{sec:placinglimits} presents the resulting limits.}
}

\appendix
\setcounter{section}{19}
\renewcommand\thefigure{\thesection.\arabic{figure}}\setcounter{figure}{0}   
\renewcommand\theequation{\thesection.\arabic{equation}}\setcounter{equation}{0}   

\subsection{GPS data processing and clock estimation}\label{sec:gpsdata}

\begin{figure*}[t]
\centering
	\includegraphics[width=0.33\textwidth]{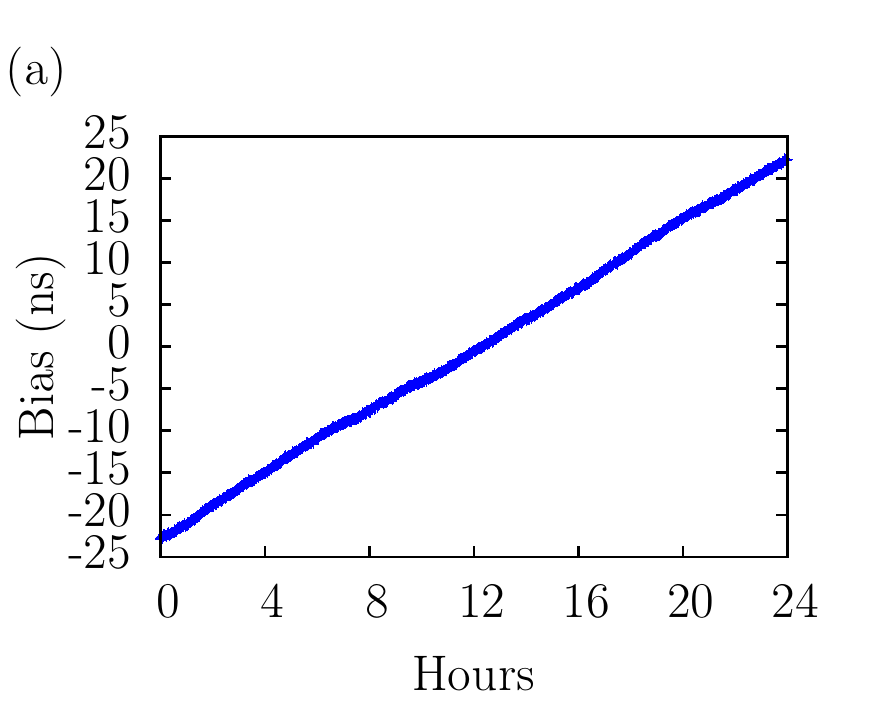}~
	\includegraphics[width=0.33\textwidth]{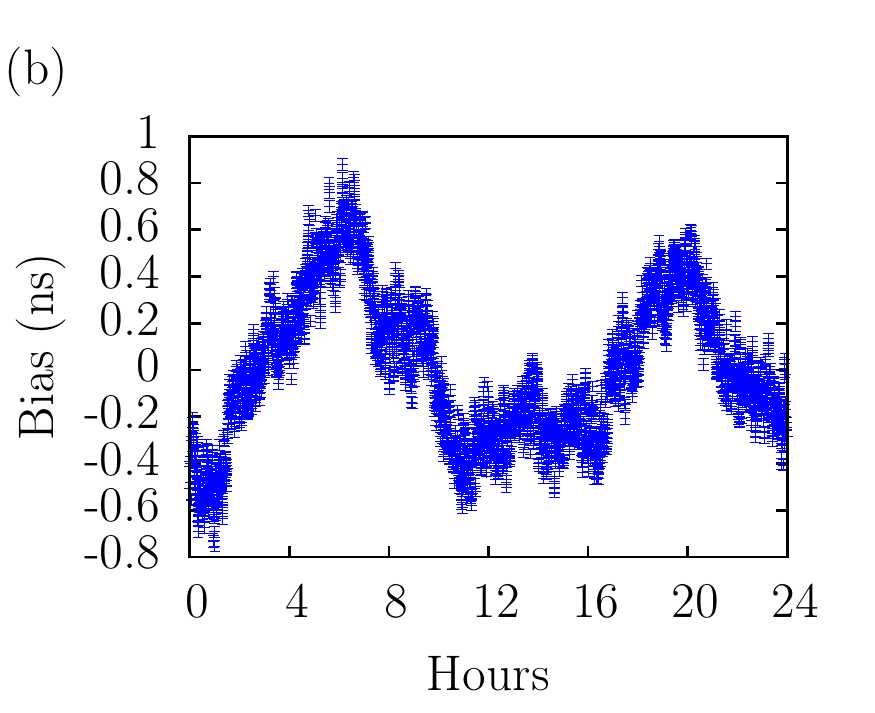}~
	\includegraphics[width=0.33\textwidth]{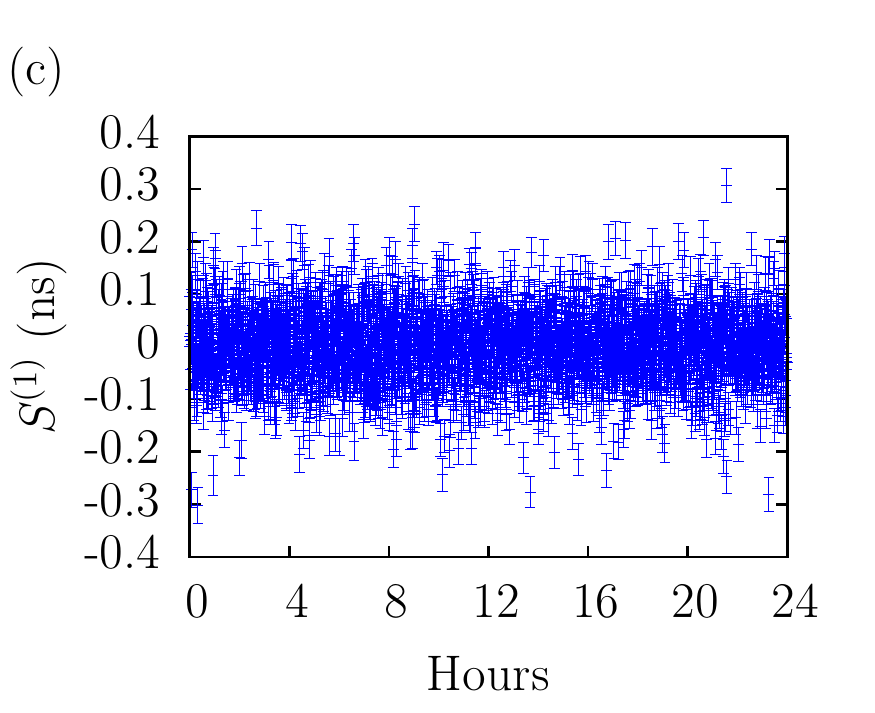}
	\caption{\small Plots of clock data from a randomly-chosen Rb clock (block IIR, Space Vehicle Number 61, see Ref.~\citenum{GPS.gov}) for 27 December 2015 (UTC); the USN7 H-maser receiver-station clock (see Ref.~\citenum{IGSnetwork}) was used as reference. {Left panel}: raw clock bias, with a constant offset removed. {Middle panel}: with a second-order polynomial fit and subtracted to reveal the sub-daily variance. {Right panel}: first-order differenced data. 
The error bars shown in the plots are the formal errors.
Note that the differenced data is generated directly from raw clock bias (no fit polynomial is removed).
}
\label{fig:data-raw-diff}
\end{figure*}

The Global Positioning System (GPS) works by broadcasting microwave signals from nominally 32 satellites in medium-Earth orbit. 
The signals are driven by an atomic clock (either based on Rb or Cs atoms) on board each satellite. 
While each satellite may host multiple Rb and Cs backup clocks, at any given time only one of these clocks drives the GPS signals, which are transmitted on carrier waves at both L1 ($1.57542\,{\rm GHz}$) and L2 ($1.2276\,{\rm GHz}$) bands.  
Superimposed on the carrier waves are streams of pseudo-random bits generated by flipping the sign of the electric field.  
A geodetic GPS receiver can sample the dual-frequency signals simultaneously from all GPS satellites in view (typically 8 to 10) at user-specified intervals (typically 1 to 30 seconds). 
At every such interval, for each satellite in view, timing measurements are made (according to the receiver clock) of the peak cross-correlation of the incoming signal with the receiver's replica model of the signal.  
At both L1 and L2 frequencies, two types of data are generated including a pseudorange (group delay, using the bits) and a carrier phase (phase delay, using the carrier wave).  
The term pseudorange is used because it is a measure of delay that is biased by the receiver clock.   
This bias cancels when differencing data between pairs of satellites.  The ionospheric delay of $\sim$\,$10\,{\rm ns}$ is calibrated with $\sim$\,$0.02\,{\rm ns}$ precision by forming a specific ``ionosphere-free'' linear combination of the data at L1 and L2 frequencies, hence the purpose of the dual-frequency system.

Measurement precision in such cross-correlation systems tends to scale with the relevant wavelength of the signal.  
In the case of pseudorange, precision scales with the time interval between bit transitions.  
As a consequence, the pseudorange precision is typically $\sim2\,{\rm ns}$, whereas in contrast, carrier phase is measured with $\sim0.02\,{\rm ns}$ precision, but suffers from a constant integer cycle bias that is initially unknown.   
Resolving this bias is known as integer ambiguity resolution. 
Combinations of the 4 observables (pseudorange and carrier phase on both frequencies) allow for robust detection of data outliers and cycle slips in the integer bias, and enable robust integer ambiguity resolution~\cite{Blewitt1989, Blewitt1990}.  
With integer ambiguities resolved, carrier phase data can then be modelled as pseudorange data, but with two orders of magnitude more precision.

Here we analyse data from the Jet Propulsion Laboratory (JPL)~\cite{JPLigsac}, in which the clock biases are given at $30\,{\rm s}$ intervals~\cite{MurphyJPL2015}. 
These clock biases are generated using data from a global network of $\sim100$ GPS geodetic receivers by a mature analysis system that is used routinely for purposes of centimetre-level satellite orbit determination, and millimetre-level positioning for scientific purposes, such as plate tectonics, Earth rotation, and geodynamics.   
The analysis standards that are applied are consistent with models and conventions specified by the International Earth Rotation and Reference Frames Service (IERS), in line with resolutions of the International Astronomical Union (IAU).  
These standards ensure best practices and consistency between various geodetic techniques, including very long baseline interferometry (VLBI), and satellite laser ranging (SLR), both of which have been instrumental in developing the IERS conventions.  
We note that similar types of analyses abiding by the IERS conventions are performed by several analysis centres around the world, and are routinely compared by the International GNSS Service. 
Results of these comparisons, along with the variety of scientific applications that depend on such data, provide an abundance of evidence that the clock biases are determined relative to each other with an accuracy $<0.1\,{\rm ns}$.  

Analysis of GPS receiver data includes the effects of special and general relativity.
All ideal clocks stationary on one of the Earth's equipotential surfaces such as the geoid (“sea level”) have no relative phase drift.  
For purposes of discussion, let us define coordinate time as proper time on the geoid. 
(Actually a different coordinate time is chosen in geodesy, but that does not change the argument here.) 
For each satellite, the combination of velocity and gravitational potential causes satellite proper time to vary with respect to coordinate time. 
This is dominated by a positive drift at the level of $0.45$ parts per billion, which would be $\sim13\,{\rm ns}$ per $30\,{\rm s}$ epoch.
The hardware in the GPS satellite is designed to set the effective frequency of the atomic clocks such that the drift is zero on average, using the known semi-major axis of each satellite's orbit.  
A residual effect results from the satellite moving in an ellipse, and thus with a time varying velocity and gravitational potential.
The eccentricity of the GPS satellites is typically small $\lesssim0.02$; nevertheless, the effect is a periodic variation of satellite proper time with an amplitude of $\sim30\,{\rm ns}$ at $\sim12\,{\rm hr}$ orbit period.  
This is modelled to first order by knowing the satellite's position and velocity, and by considering the Earth as a spherically symmetric mass. 
All higher order effects effectively go into the definition of the satellite clock bias provided by JPL, consistent with international conventions. 
These residual effects are less than $1\,{\rm ns}$ over the $\sim12\,{\rm hr}$ orbit period, and so are completely negligible over the time periods investigated here.

In JPL's global GPS data analysis, the results of which are employed as input data to our dark matter (DM) search, the clock biases are estimated using GPS ionosphere-free carrier phase and pseudorange data combinations, as part of a multi-parameter least-squares estimation process using a square-root information filter.   
Other parameters that are estimated along with the clock biases include GPS orbits, station positions, Earth rotation, and atmospheric delay. 
Correlated errors between the clocks and other parameters are formally computed by the square-root information filter to be at the level 
$< 0.1\,{\rm ns}$, consistent with the inferred level of accuracy. 

In this process, there is effectively no restriction on the allowed behaviour of the clocks from one $30\,{\rm s}$ epoch to the next.
Crucially, if a clock were to have a real transient that far exceeded engineering expectations, the data over that time window would not have been removed as outliers. 
Since only relative clock bias can be estimated, one clock is always held fixed in the estimation procedure. 
The choice of reference clock is irrelevant, in that our search algorithms look at differences in estimated clock biases.

In order to analyse the estimated clock biases, we further apply a first-order differencing procedure to the data, and define the pseudo-frequencies
\begin{equation}
S^{(1)}(t_k)=S^{(0)}(t_k)-S^{(0)}({t_{k-1}}),
\label{eq:S1-diff}
\end{equation}
where $S^{(0)}(t_k)$ is the original clock bias for the $t_k$ epoch (data point), and $t_{k}-t_{k-1}=30\,{\rm s}$ is the sampling interval. 
This is equivalent to taking a discrete derivative (up to a multiplicative factor), and acts to whiten the data (since the clock bias noise is dominated by random walk processes).
This differencing procedure also removes any constant bias offsets, and transforms the linear frequency drifts into constant offsets in $S^{(1)}$. In practice, these residual offsets are small, and are removed by subtracting the mean of $S^{(1)}$ over the given day.
The effect of this procedure is demonstrated for a single arbitrarily-chosen clock in  Fig.~\ref{fig:data-raw-diff}.

\begin{figure}
\centering
	\includegraphics[width=0.475\textwidth]{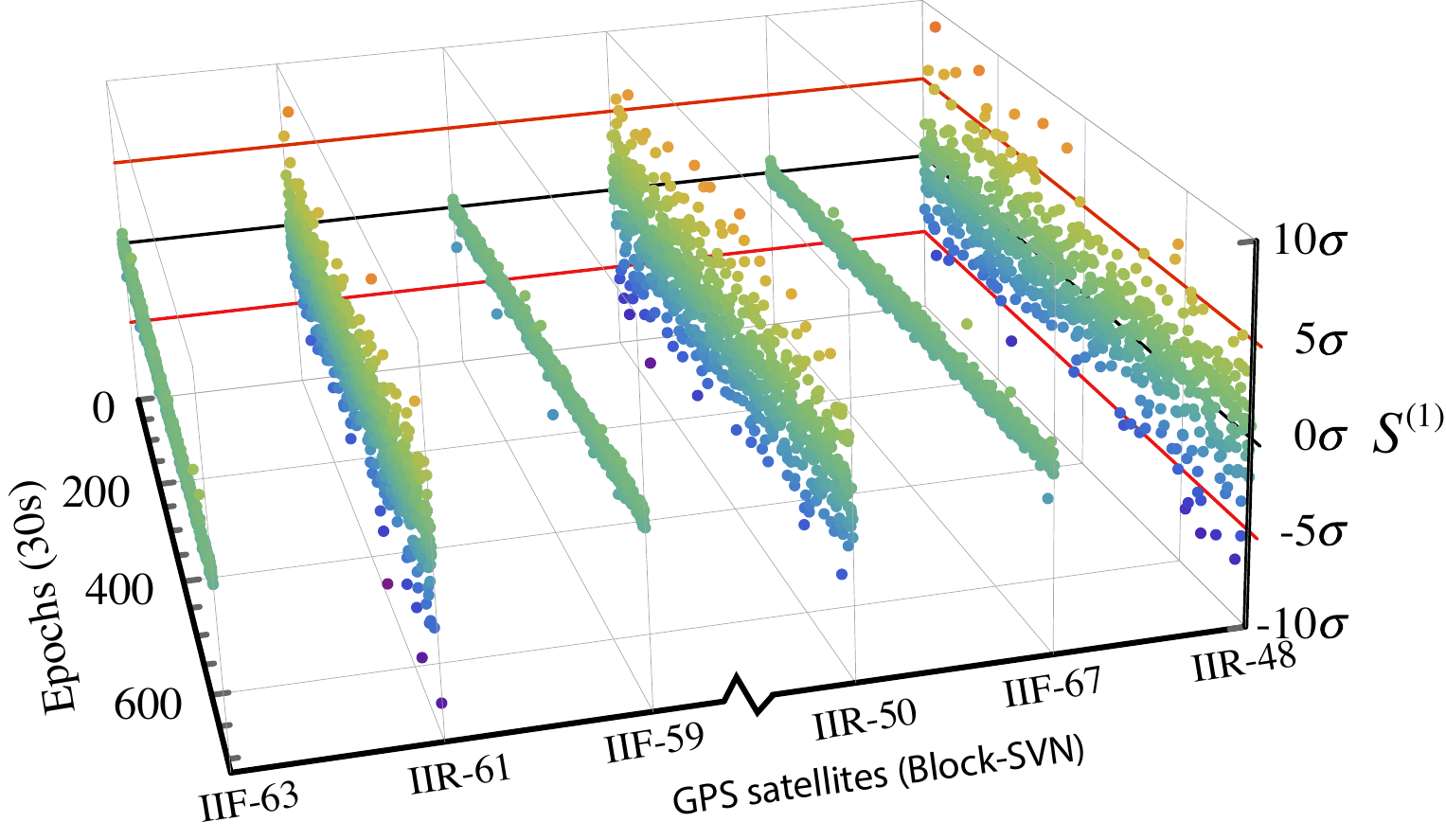} 
	\caption{\small The first 6 hours of single-differenced clock data $S^{(1)}$, in units of the
formal error $\sigma$, for 5 June 2016 (UTC), for six arbitrarily chosen Rb clocks.
The clocks are labelled by their satellite block (either II, IIA, IIR or IIF), and their Space Vehicle Number (SVN); see Ref.~\citenum{GPS.gov}. 
}
\label{fig:waterfall}
\end{figure}

The clock biases from JPL~\cite{JPLigsac} also come with a ``formal error''.
The formal error quantifies uncertainty in the determination of the clock bias, and does not directly incorporate the intrinsic clock noise.
It varies slowly over time, and is typically on the order of $\sim0.02-0.03\,{\rm ns}$; the formal error is dominated by the uncertainty in the satellite orbit determination.
Only the most recent satellite clocks have observed temporal variations in $S^{(1)}$ that are at a similar level as the formal error, indicating that temporal variations in older clocks are actually due to clock behaviour rather than estimation error.  
A snapshot of the standard deviations for the $S^{(1)}$ clock data used in our analysis is presented in Table~\ref{tab:sd}.
A plot showing a few hours of $S^{(1)}$ data for an arbitrarily selected few satellite clocks is shown in Fig.~\ref{fig:waterfall}.
This shows quite clearly how the more modern block IIF satellites~\cite{GPS.gov}  are substantially less noisy than the older-generations of satellites (by orders of magnitude).

\begin{table} 
\centering
\caption{\small Typical standard deviations (in ns) of the $S^{(1)}$ clock data over the 16 years analysed in this work. The standard deviations vary significantly between the satellite generations~\cite{GPS.gov} (II, IIA, IIR, and IIF), and also depend on whether the ground based H-maser clock was used as reference (in this case USN3~\cite{IGSnetwork}), or one of the other GPS satellite clocks.
}
\begin{tabular}{lll}
\hline
GPS Block &\multicolumn{2}{c}{Reference}\\
\cline{2-3}
&USN3&RbIIR\\
\hline
RbII&0.047&0.070\\
RbIIA&0.038&0.074\\
RbIIR&0.073&0.097\\
RbIIF&0.013&0.067\\
\hline
GPS Block &\multicolumn{2}{c}{Reference}\\
\cline{2-3}
 &USN3&CsIIA\\
\hline
CsII&0.081&0.112\\
CsIIA&0.088&0.124\\
CsIIF&0.098&0.128\\
\hline
\end{tabular}
\label{tab:sd}
\end{table}

Since the clocks are noisy, in order to discern the DM-induced signal from the intrinsic clock noise we rely on signals correlated across the entire GPS network, as discussed in the following sections.
Moreover, there already exists more than 15 years of high-accuracy timing data that can be exploited in the search. 
This data stream is being routinely updated, and, in principal, timing data from any other atomic clocks that are synchronised with GPS can be included in the analysis (for a discussion on this point, see Ref.~\citenum{BudkerPT2015}).
Therefore, by analysing the new and existing GPS timing data, we can perform a sensitive search for transient DM signals, and if no signals are found we can then place limits on the DM--ordinary matter interaction strengths \cite{DereviankoDM2014}.
The GPS network is particularly well suited for this type of search, for a number of reasons.
The large number of clocks, and the very large ($\sim$\,$50\,000\,{\rm km}$) diameter of the network increases both the chance of an interaction, and the sensitivity of the search, since we rely on correlated signal throughout the entire network.
Similar arguments underpin motivations for searches using global network of atomic magnetometers~\cite{Pospelov2013,Pustelny2013}. 
Such magnetometry searches are sensitive to different types of interactions (as discussed below), and are therefore complementary to atomic clock searches.

\subsection{Using GPS to search for topological defects}\label{sec:GPSTDSM}

The null results from recent  WIMP (weakly-interacting massive particle) direct-detection experiments have partly contributed to increased attention to ultralight field DM, such as axions \cite{Peccei1977a,Peccei1977b,Srednicki1981,Sikivie1983,Preskill1983}.
Ultralight fields may form stable topological defects (TDs), such as monopoles, strings, or domain walls, which can be a dominant or subdominant contribution to both DM and dark energy
\cite{
Sikivie1982,
Press1989,
Vilenkin1994,
Battye1999,
Durrer2002,
Friedland2003,
Avelino2008%
}.
The interactions of light fields with standard model (SM) fields can be written as a sum of effective interaction Lagrangians (so-called ``portals'') \cite{DereviankoDM2014}
\begin{equation}\label{eq:effL}
\mathcal{L}_{\rm int} = \mathcal{L}^{\rm PS} + \mathcal{L}^{\rm S^1} +\mathcal{L}^{\rm S^2}  + \; \ldots \; ,
\end{equation}
where $\mathcal{L}^{\rm PS}$ represents the pseudoscalar (axionic) portal, and $\mathcal{L}^{\rm S^1}$ and $\mathcal{L}^{\rm S^2}$ are the linear and quadratic scalar portals, respectively.
The linear and quadratic scalar portals, as will be demonstrated below, lead to changes in the effective values of certain fundamental constants and thus cause shifts in atomic transition frequencies, and so lend themselves well to searches based on the use of atomic clocks.
The axionic portal leads to interactions that could cause spin-dependent shifts due to fictitious magnetic fields, and are thus are well suited to magnetometry searches \cite{Pospelov2013,Pustelny2013}.
We note that there are very stringent limits on the interaction strength for the linear scalar interaction coming from astrophysics and gravitational experiments (see, e.g., Refs.~\citenum{Raffelt1999,Bertotti2003}).
However, the constraints on the quadratic portal are substantially weaker \cite{Olive2008}. 
For this reason, we will concentrate on the quadratic scalar portal.

While we mainly address topological defect dark matter, and in particular refer the quadratic scalar coupling, it is important to note that our experiment is not limited in scope to this possibility.
Any large (on laboratory scales), ``clumpy'' object (e.g.,~$Q$-balls~\cite{Coleman1985,Dvali1998,Kusenko2001}, strings~\cite{Hogan1984}, solitons~\cite{Marsh2015,Kusenko1997,Lee1992}, and other stable objects~\cite{Hogan1988,Kolb1993,Jetzer1992}) that interacts with standard model particles in such a way that leads to shifts in atomic transition frequencies is possible to detect using this scheme.

In the assumption of a quadratic scalar coupling between the standard model (SM) and DM fields, the interaction Lagrangian can be parameterized as \cite{DereviankoDM2014}
\[
-\mathcal{L}^{\rm DM-SM}=  \qquad\qquad\qquad\qquad\qquad\qquad
\]
\begin{equation}
\label{eq:scalarPortal}
\varphi^2 \left( \mathbf{r},t \right)
\left(
{\Gamma_f}\,{m_f c^2 \bar \psi_f \psi_f  }
 +\frac{\Gamma_\alpha}{4}F_{\mu\nu}F^{\mu\nu} 
+ \,\ldots \right),
\end{equation}
where $\varphi$ is the DM field, $m_f$ are the fermion masses, $\psi_f$ and $F_{\mu\nu}$ are the SM fermion fields and electromagnetic field tensor, respectively, and there is an implicit sum over $f$ that runs over all SM fermions. 
The constants $\Gamma$ quantify the strengths of the various DM--SM couplings.
From a comparison with the conventional SM Lagrangian, 
\[
-\mathcal{L}^{\rm SM} = m_f c^2\,\bar \psi_f \psi_f  +\frac{1}{4}F_{\mu\nu}F^{\mu\nu}    
+ \,\ldots \,,
\]
it is seen that (to lowest order) the above Lagrangian (\ref{eq:scalarPortal}) leads to the effective redefinition of certain dimensionless combinations of fundamental constants:
\begin{equation}
\label{eq:varalpha}
\alpha^{\rm eff} = {\alpha }\left(1+ \Gamma_{\alpha}\,{\varphi^2}\right)\; ,
\end{equation}
\begin{equation}
\label{eq:mep}
m_{e,p}^{\rm eff} = {m_{e,p}}\left(1+ \Gamma_{m_{e,p}}\,{\varphi^2}\right)\; ,
\end{equation}
\begin{equation}
\left(\frac{m_q}{\Lambda_{\rm QCD}}\right)^{\rm eff} = \frac{m_{q}}{\Lambda_{\rm QCD}}\left(1+ \Gamma_{{m_{q}}/{\Lambda_{\rm QCD}}}\,{\varphi^2}\right),
\label{eq:varqcd}
\end{equation}
where $\alpha$ and $\Lambda_{\rm QCD}$ are the nominal values of the fine structure constant and the QCD energy scale, respectively,
and $m_{e,p}$ are the electron and proton masses. 
In our notation, the DM field has units of energy $(E)$, thereby $\Gamma_X$ is expressed in units of $E^{-2}$.
Following Ref.~\citenum{DereviankoDM2014}, and to aid in the comparison with other works, we will present our results in terms of effective energy scales 
 $\Lambda_X=1/\sqrt{|{\Gamma_X}|}$ (for $X=\alpha,\,m_e,\,m_p,\,m_q/\Lambda_{\rm QCD}$).
Note that, by the nature of topological defects (TDs), the DM field $\varphi^2\to0$ outside the defect, and $\varphi^2\to \varphi_{\rm max}^2$ inside the defect; as such the redefined coupling constants are only realised inside the defect.
The field amplitude, $\varphi_{\rm max}$, can be linked to $\rho_{\rm inside}$, the energy density inside the defect, as 
\begin{equation}
\rho_{\rm inside}= \varphi_{\rm max}^2/(\hbar c\, d^2),
\end{equation}
 where $d$ is the width of the defect, which is set by the Compton wavelength for the field \cite{DereviankoDM2014}:
\begin{equation}\label{eq:compton}
d=\hbar/(m_\varphi c).
\end{equation}
Note the ultra-light mass scale for the fields we are interested in: a roughly Earth-sized defect has a mass scale $m_\varphi\sim10^{-14}{\,\rm eV}$ ($m_\varphi$ refers to the mass of the field particles, not the mass of the defect itself).
We note here that there are other possibilities to search for ultralight DM fields via their non-gravitational interactions; see, e.g., 
 Refs.~\citenum{Pospelov2013,Pustelny2013,
Budker2014,
StadnikDefects2014,
Arvanitaki2014,Arvanitaki2015,
StadnikDMalpha2015,Arvanitaki2014b,
StadnikLasInf2015,StadnikLaser2015,Arvanitaki2016,
Hall2016}.

From Eqs.~(\ref{eq:varalpha}) -- (\ref{eq:varqcd}), we may relate the DM-induced frequency shift to a transient variation of fundamental constants.
The fractional shift in the clock atom transition frequency, $\omega_c$, can be expressed as
\begin{equation}
 \frac{\delta \omega(t)}{ \omega_c} = \sum_X K_X \frac{ \delta X(t)}{X} 
		\equiv \Gamma_{\rm eff}\,{\varphi^2}
  ,  \label{eq:variation}
\end{equation}
where $K_X$ are dimensionless sensitivity coefficients, and $X$ runs over fundamental constants from (\ref{eq:varalpha})--(\ref{eq:mep}), 
\begin{equation}
\frac{ \delta X(t)}{X} = \Gamma_X{ \varphi^2},
\label{eq:dXonX}
\end{equation}
and the overall effective coupling constant is defined as
\begin{equation}\Gamma_{\rm eff}\equiv \sum_X K_X\Gamma_X, \end{equation}
which depends on the specific clock transition through the sensitivity coefficients $K_X$.
We also define $\Lambda_{\rm eff} = 1/\sqrt{|\Gamma_{\rm eff}|}$, which have a meaning of effective energy scales.
The dimensionless sensitivity coefficients $K_X$ are known from atomic and nuclear structure calculations~\cite{Dzuba2003}.
For example, considering only the variation in the fine structure constant, $\alpha$, and ignoring relativistic effects,
for optical and microwave transitions, the clock frequencies scale as
$\omega^{\rm opt}_c \propto \alpha^2$, and
$\omega^{\rm mw}_c \propto \alpha^4$, respectively.
Slight deviations from these dependencies occur due to relativistic corrections to the atomic structure~\cite{Dzuba2003,Angstmann2004}.

The clocks on board the GPS satellites work by disciplining the frequency of a quartz oscillator to the resonant frequency of the chosen atomic transition; the quartz oscillator drives the GPS L1 and L2 microwave signals \cite{Dupuis2008}.
In order to make a meaningful analysis of the data we must analyse the comparison of the output frequencies to the frequency of another reference clock.
For thin domain walls, the GPS clock and the reference clock are spatially separated, so only one is affected by the TD at a given time.
(When both clocks are affected at the same time we would see no DM signal, as discussed in the next section.)
Therefore, for the microwave frequency $^{87}$Rb and $^{133}$Cs clocks 
on board the GPS satellites,
the sensitivity coefficients are \cite{Flambaum2006a,Dinh2009}
\begin{equation}
\label{eq:KRb}
\frac{\delta \omega}{\omega_c}({\rm Rb})
=\varphi^2({4.34}\,{\Gamma_\alpha}-{0.019}\,{\Gamma_q}+{\Gamma_{e/p}}), 
\end{equation}
\begin{equation}
\label{eq:KCs}
\frac{\delta \omega}{\omega_c}({\rm Cs})
=\varphi^2({4.83}\,{\Gamma_\alpha}+{0.002}\,{\Gamma_q}+{\Gamma_{e/p}}), 
\end{equation}
respectively, where we have defined the short-hand notation $\Gamma_{e/p}\equiv2\Gamma_{m_e}-\Gamma_{m_p}$, and $\Gamma_q\equiv\Gamma_{m_q/\Lambda_{\rm QCD}}$ (and similarly for $\Lambda_{e/p}$ and $\Lambda_q$). 
Note that the value of $K_q$ comes from a combination of a shift in the nuclear magnetic moment, $K_\mu$ and a shift in the nuclear size, $K_{\rm hq}$.
For Cs, the contributions from these two effects are roughly equal in magnitude and opposite in sign~\cite{Dinh2009} ($K_q^{\rm Cs}=K_\mu+K_{\rm hq}=0.009-0.007$), and $K_q$ in Cs is therefore sensitive to uncertainties in the nuclear calculations.
For clocks based on optical transitions, there is only sensitivity to variations in the fine structure constant~\cite{Dzuba2003,Angstmann2004}.
For clocks that compare the clock frequency to the resonant frequency of an optical cavity, there is an extra factor of $\alpha$ that comes from the variation in the length of the cavity, e.g.,
\begin{equation}
\label{eq:KSr}
\frac{\delta \omega}{\omega_c}({\rm Sr_{\rm optical}})=\varphi^2\,{1.06}\,{\Gamma_\alpha},   
\end{equation}
as in Ref.~\citenum{Wcislo2016}.  
In general, the cavity contributes no extra sensitivity to variations in $m_e/m_p$ or $q/\Lambda_{\rm QCD}$ \cite{StadnikLaser2015,StadnikLasInf2015}.

The fact that there are a number of different atomic clock types employed within the GPS network means we have sensitivity to several different combinations of the parameters (\ref{eq:variation}).
The existing limits come from constraints on supernova emission \cite{Olive2008}:
{${\Lambda_{m_e,\alpha}\gtrsim3\,{\rm TeV}}$}, and 
$\Lambda_{m_p}\gtrsim15\,{\rm TeV}$, with no existing limit on $\Lambda_q$.

\subsection{Domain wall signal}\label{sec:DomainWall}

One specific example of a TD is a domain wall, a quasi-2D structure that can be characterised by a width, $d$, and DM field amplitude~\cite{DereviankoDM2014}, $\varphi_{\rm max}$ (see Fig.~1 of main text).
We assume a Gaussian distribution of the field across the wall.
As a wall passes a clock, it causes a shift in the clock frequency $\omega_c\to\omega_c+\delta\omega(t)$, where $\delta\omega(t)$ 
is proportional to the square of the field value and rapidly goes to zero outside of the wall.
Then, the time-difference due to a transient frequency shift with respect to an unperturbed clock is
$\Delta t= \int_{-\infty}^t \frac{\delta \omega(t')}{\omega_c} dt'$.
The accumulated time-reading bias of clock $i$ (measured against some reference clock, $R$) at time $t$ caused by a Gaussian-profile domain-wall that crosses the clock at time $t^\times_i$ and the reference clock at $t^\times_R$, is
\[
S^{(0)}_i(t) = \varphi_{\rm max}^2\int_{-\infty}^t  \Bigg[{\Gamma_{{\rm eff,}i}}\,\exp\left(\frac{-(t'-t^\times_i)^2}{\tau^2}\right)\qquad\qquad
\]
\begin{equation}
\qquad\qquad- {\Gamma_{{\rm eff,}R}}\,\exp\left(\frac{-(t'-t^\times_R)^2}{\tau^2}\right)\Bigg]dt',
\label{eq:Sit}
\end{equation}
where the (single-clock) crossing duration is defined as
\begin{equation}\label{eq:tau}
\tau\equiv\frac{d}{v_\perp},
\end{equation}
which is the time scale during which the clock is inside the wall, and $v_\perp$ is the component of the wall's velocity that is perpendicular to the wall.

\begin{figure}
\centering
	\includegraphics[width=0.42\textwidth]{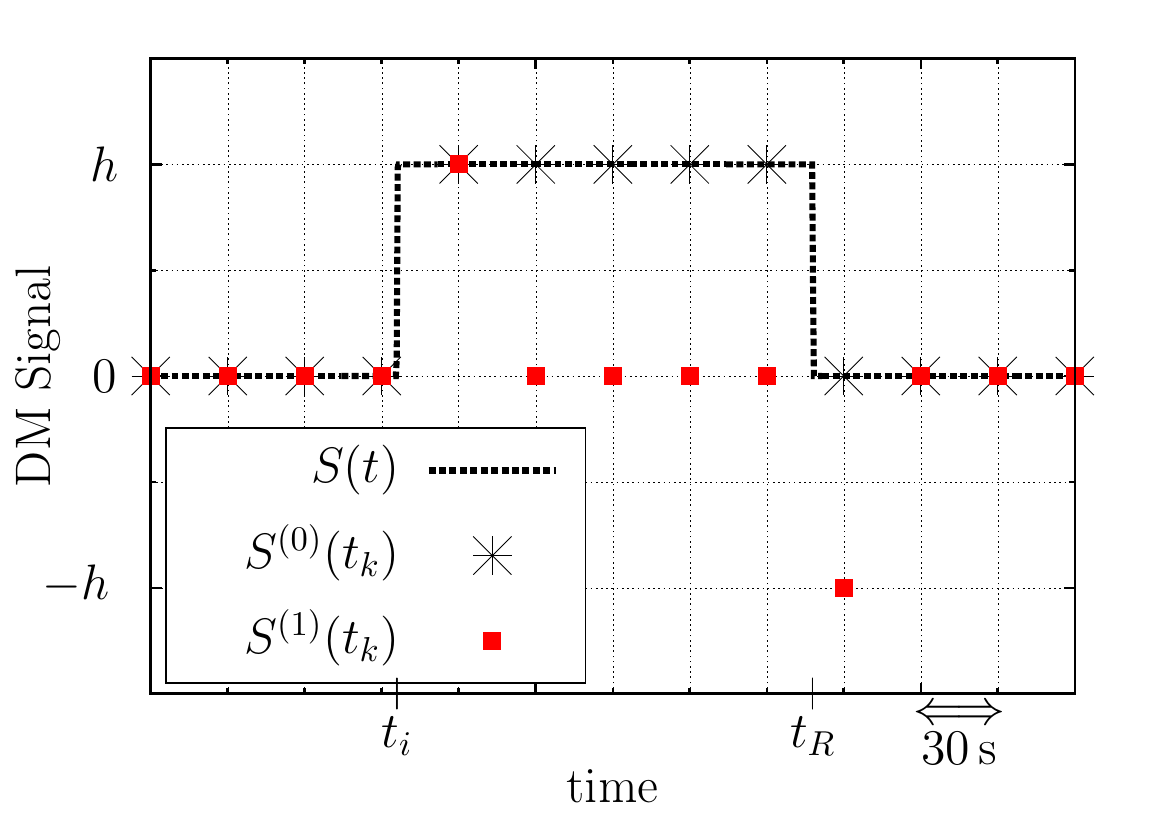}
	\caption{\small Example of an ideal thin-wall signal for a pair of identical clocks separated by a distance $l=v(t_R-t_i)$, where the wall crosses the first clock at time $t_i$, and the second (reference) clock at time $t_R$. 
As the wall passes the first clock, it causes a shift in the clock frequency $\omega_c\to\omega_c+\delta\omega(t)$, where $\delta\omega(t)$ is only non-zero while the wall encompasses the clock. This leads to a difference in the time-readings (bias) between the two clocks, with magnitude $h\approx \delta\omega(t_i)\,\tau$, where $\tau=d/v_\perp$ is the interaction time. When the wall crosses the second clock, it causes the same frequency shift, so that after the wall has crossed, the time-reading difference returns to zero.
	The magnitude of the time-difference depends on the DM--SM coupling strength, the speed of the wall (i.e.~the duration of the interaction), and the width of the wall.
The red squares correspond to the signal after applying the differencing technique, Eq.~(\ref{eq:S1-diff}).}
\label{fig:S1}
\end{figure}

As mentioned above,  to analyse the data, we first perform a single-differencing procedure (\ref{eq:S1-diff}).
We perform the same differencing procedure for the DM signals.
The correspondence between the $S^{(0)}$ and $S^{(1)}$ signals is shown in Fig.~\ref{fig:S1}.

For thin walls (that is, for walls thin enough such that the interaction time is smaller than the data acquisition interval, $\tau<30\,{\rm s}$), the $S^{(1)}$ spike amplitude would see in the first-order differenced data at the time of the wall crossing (for a system of identical clocks) is
\begin{equation}\label{eq:S1form}
S^{(1)} = {\varphi_{\rm max}^2}\,{\Gamma_{\rm eff}}\,\sqrt{\pi}\,\tau,
\end{equation}
as shown in Fig.~\ref{fig:S1}.
Note that in the ``thin'' wall case, the assumed Gaussian profile is unimportant; other profiles (e.g., hard spheres, flat profiles) give similar results.
These non-Gaussian profiles can be incorporated through the form factors in Eq.~(\ref{eq:S1form}) arising from the integrals in Eq.~(\ref{eq:Sit}).
Our following analysis holds equally in these cases, the difference being only the ratio of the specific form factors.
For example, for a flat (hard-edge) profile, $\Gamma_{\rm eff}\to\Gamma_{\rm eff}/\sqrt{\pi}$ in Eq.~(\ref{eq:S1form}).

In order to determine the expected average $S^{(1)}$, we use the average value for the single-clock crossing duration, $\tau_{\rm avg}$, which depends on the object width $d$, orientation, and velocity distribution, as will be discussed in Section~\ref{sec:sensitivity}.
Further, in the assumption that the TDs saturate the DM energy density in the galaxy, one can link $\varphi_{\rm max}$ and $d$ to $\rho_{\rm DM}$:
\begin{equation}
\varphi_{\rm max}^2 = \hbar c \rho_{\rm DM} d^2 \frac{\mathcal{T}}{\tau_{\rm avg}},
\end{equation}
where 
$\mathcal{T}$
is the average time between events (i.e., encounters between the GPS constellation and DM objects).
Therefore, the average DM signal amplitude is:
\begin{equation}
\label{eq:S1}
S^{(1)}_{\rm avg}  = {\hbar c}\,{\Gamma_{\rm eff}} \, \sqrt{\pi} \rho_{\rm DM}\, d^2 \,\mathcal{T}.
\end{equation}
This equation translates into constraints on $\Gamma_{\rm eff}$, Eq.~(5) of the main text.

\subsection{Search}\label{sec:window}

Here, we perform a search with the aim of ruling out or detecting large ($\gtrsim5\sigma$) events
(here, $\sigma$ is the standard deviation of $S^{(1)}$ clock data, which is about $0.1\,{\rm ns}$ in the worst-case for the Rb clock solutions when using another Rb clock as a reference, see Table~\ref{tab:sd}).
Consider a time-period window, denoted $J_w$.
If a thin-wall TD-DM object passes through the GPS constellation with a speed 
\begin{equation}\label{eq:Jw}
v>D_{\rm GPS}/J_w,
\end{equation}
where $D_{\rm GPS}\approx50\,000\,{\rm km}$ is the GPS orbital diameter, we can expect that all of the clocks will be affected by the object within this window.
However, any clock that is swept within the same 30\,{\rm s} sampling interval as the reference clock will not show any spike in the $S^{(1)}$ data due to clock degeneracy (see Fig.~\ref{fig:S1}, or satellites 15 and 16 in Fig.~3(a) of the main text).
For a DM object with relative velocity
$v<700\,{\rm km\,s}^{-1},$
we can safely expect at least 60\% of the clocks to show an $S^{(1)}$ spike (this number comes from the orbital geometry of the GPS constellation, noting that by design the satellites are relatively evenly distributed over the sky).
For more typical velocities, $v\sim100-300\,{\rm km\,s}^{-1}$, more than 80\% of the satellites are expected to show the $S^{(1)}$ spikes.

To ensure that the potential DM-induced frequency excursions for each clock are of the same magnitude, we only consider sub-networks of identical clocks.
For example, for the Rb sub-network, we choose the most recently launched Rb satellite to be the reference clock (as a rule such clocks are the least noisy), and compare all the other Rb clocks to this reference, and do not include the Cs or H-maser clocks in the analysis.
To unfold limits on the various coupling constants in Eqs.~(\ref{eq:KRb}) and (\ref{eq:KCs}), we only consider the Rb and Cs sub-networks, and supplement them with the Sr limits on $\Gamma_\alpha$ from Ref.~\citenum{Wcislo2016}. 
We do not include the ground clocks in our current analysis.

To search for domain wall signals, we analysed the $S^{(1)}$ GPS data streams in two stages. The first stage involved stepping through all the data, one epoch at a time, to identify regions of interest that could potentially be consistent with thin-wall crossings.
We call these regions ``potential events''.
The second stage investigates these regions using a more detailed approach, to determine if we can upgrade any potential events to ``candidate events''.

At the first stage, we scanned all the data from May 2000 to October 2016 looking for the most general patterns associated with a domain wall crossing, without taking into account the order in which the satellites were swept.
We required at least 60\% of the clocks to experience a frequency excursion at the same epoch, as discussed above. 
This procedure identifies the epoch 
when the wall crossed the reference clock (vertical blue line in Fig.~3(a) of the main text).
If such a frequency excursion is found at epoch $t_0$, we form all the possible windows, $J_w$, around $t_0$, in the range $3\leq J_w/30\,{\rm s}\leq500$.
The chosen maximum $J_w$ determines the minimum DM velocity we can detect (\ref{eq:Jw}). 
Our choice of a maximum window of $500\,{\rm epochs}$, corresponding to sweep durations through the GPS constellation of up to 15,000\,{\rm s}, ensures sensitivity to walls moving as slowly as $\sim 4\,{\rm km\,s}^{-1}$. 
Less than 0.1\% of DM domain walls that cross the GPS network are expected to travel with perpendicular velocities slower than this value.

We define a ``potential event'' as any time where at least the given fraction of the clocks also experience a frequency excursion of sign opposite to the reference clock excursion anywhere within that window (red tiles in Fig.~3 of main text). 
Since we consider only the sub-networks of identical clocks (e.g., we only consider the Rb clocks), if an event occurs within the window all the DM-induced frequency excursions must be of the same magnitude (besides the clock noise).
Therefore, we only search for frequency excursions within a range of $S^{(1)}$ values, 
$S^{(1)}_{\rm cut}\to S^{(1)}_{\rm cut}\pm dS$, where $dS=x\sigma$, 
with $\sigma$ being the calculated standard deviation for that given GPS clock/reference clock combination (see Table~\ref{tab:sd}); for 90\% confidence level (C.L.), $x=1.645$.
(By using the calculated standard deviation here, we account for both the formal error (see section~\ref{sec:gpsdata}) and the clock noise directly.)
We count the total number of potential events for each window-size and $S^{(1)}_{\rm cut}$ combination.

This method also allows us to place more stringent limits on events for which the average time between events, $\mathcal{T}$, is less than the observation time, $\mathcal{T}_{\rm obs}$.
For example, if for a given $S^{(1)}_{\rm cut}$, we saw 10 potential events in a 10 year observation time, we can place a limit at this level for $\mathcal{T}\simeq1\,{\rm yr}$.
Further, we only include days in our analysis for which there is data available for at least 7 clocks; this effectively reduced our observation time, especially for the Cs sub-network, which employs only a small number of clocks in recent times.
Note, the choice of 7 simultaneously operational clocks here is conservative; 
injecting fake DM events demonstrates that this search technique would find practically every event (that otherwise falls within our current sensitivity) with at least 5 available clocks.

The tiled representation of the GPS data stream depends on the chosen signal cut-off $S^{(1)}_\mathrm{cut}$ (see Fig.~\ref{fig:waterfall} and Fig.~3 of the main text). 
We systematically decreased the cut-off values and repeated the above analysis.
Wherever the counted number of ``potential events'' goes to zero gives the overall limit for the total observation time, which we denote $S^{(1)}_{\rm thresh}$. 
For example, for a window size encompassing sweeps at $v\sim300\,{\rm km\,s}^{-1}$ ($J_w=170\,{\rm s}$), we can exclude events in the Rb network above the $S^{(1)}_{\rm thresh}=0.48\,{\rm ns}$ level (90\% C.L.), for a total observation time of $\mathcal{T}_{\rm obs}=16.5\,{\rm yrs}$.
Below the thresholds, a number of potential events were identified.
An illistrative subset of the results of this analysis is presented in Table~\ref{tab:S1-2-Rb} for the Rb sub-network, and in Table~\ref{tab:S1-2-Cs} for the Cs sub-network.

The second stage of the search involved analysing the ``potential events'' in more detail, so that we may elevate their status to ``candidate events'' if warranted by the evidence.
We examined a few hundred potential events that had $S^{(1)}$ magnitudes just below $S^{(1)}_\mathrm{thresh}$, by matching the data streams against the expected patterns; one such example is shown in Fig.~3(a).
At this second stage, we accounted for the ordering and time at which each satellite clock was affected.
The velocity vector and wall orientation were treated as free parameters within the bounds of the standard halo model, and we only considered time window lengths of $J_w<50\,{\rm epochs}$ ($v_{\rm min}\approx30\,{\rm km\,s}^{-1}$).
As a result of this pattern matching, we did not find any events that were consistent with domain wall DM, thus we have found no candidate events at our current sensitivity.
Analyzing numerous potential events well below $S^{(1)}_{\rm thresh}$ (see Tables~\ref{tab:S1-2-Rb} and \ref{tab:S1-2-Cs}) has proven to be substantially more computationally demanding, and is beyond the scope of the current work.

While the detailed second stage analysis have improved (lowered) the values of $S^{(1)}_{\rm thresh}$ by rejecting a few hundred potential events, the improvement was minor.
Therefore, to be conservative, we have used the $S^{(1)}_{\rm thresh}$ results from the first stage of the analysis to place our constraints on DM walls.

\begin{table*}
\centering
\caption{\small An illistrative subset of the results of the stage 1 analysis, showing the 
number of potential events for the given cut-off values $S^{(1)}_{\rm cut}$ (using a 90\% C.L. range) for the Rb satellite GPS sub-network for various time window lengths $J_w$. 
$S^{(1)}_{\rm thresh}$ is defined as the smallest $S^{(1)}_{\rm cut}$ value with zero potential events.
The analysis covers a total effective period of 16.5 years, and includes a total of 131\,353 clock-days of data.
}
\begin{tabular}{l|rrrrrrr}
\hline
& \multicolumn{7}{c}{$J_w$\,(s) }\\
$S^{(1)}_{\rm cut}$\,(ns)	&	90	&	270	&	450	&	630	&	810	&	990	&	1170\\
\hline
0.1	&	4295864	&	4387849	&	4387849	&	4387849	&	4387849	&	4387849	&	4387849\\
0.15	&	682150	&	973116	&	973133	&	973133	&	973133	&	973133	&	973133\\
0.2	&	69619	&	210761	&	211120	&	211127	&	211127	&	211127	&	211127\\
0.25	&	4969	&	46234	&	47214	&	47310	&	47322	&	47328	&	47330\\
0.3	&	363	&	10233	&	11727	&	11906	&	11955	&	11974	&	11994\\
0.35	&	23	&	1331	&	2290	&	2594	&	2680	&	2724	&	2737\\
0.4	&	1	&	97	&	235	&	345	&	385	&	413	&	437\\
0.45	&	0	&	5	&	20	&	29	&	39	&	45	&	51\\
0.5	&	0	&	1	&	3	&	4	&	4	&	5	&	5\\
0.55	&	0	&	0	&	1	&	1	&	1	&	1	&	1\\
0.6	&	0	&	0	&	0	&	1	&	1	&	1	&	1\\
0.65	&	0	&	0	&	0	&	0	&	0	&	0	&	0\\
\hline
\end{tabular}
\label{tab:S1-2-Rb}
\end{table*}

\begin{table*}
\centering
\caption{\small 
As in Table~\ref{tab:S1-2-Rb}, but for the Cs GPS sub-network.
 Here, the analysis covers a total effective period of 10.5 years, and includes a total of 37\,099 clock-days of data.
}
\begin{tabular}{l|rrrrrrr}
\hline
& \multicolumn{7}{c}{$J_w$\,(s) }\\
$S^{(1)}_{\rm cut}$\,(ns)	&	90	&	270	&	450	&	630	&	810	&	990	&	1170\\
\hline
0.1	&	6807411	&	6885078	&	6885078	&	6885078	&	6885078	&	6885078	&	6885078\\
0.15	&	2283645	&	3123230	&	3123326	&	3123326	&	3123326	&	3123326	&	3123326\\
0.2	&	476928	&	1490303	&	1492203	&	1492225	&	1492225	&	1492225	&	1492225\\
0.25	&	60719	&	552654	&	566282	&	566538	&	566555	&	566559	&	566559\\
0.3	&	6525	&	148529	&	171379	&	173184	&	173369	&	173395	&	173402\\
0.35	&	512	&	28240	&	40620	&	43181	&	43822	&	43993	&	44050\\
0.4	&	39	&	4264	&	7660	&	8982	&	9513	&	9742	&	9845\\
0.45	&	3	&	495	&	1087	&	1442	&	1655	&	1777	&	1883\\
0.5	&	0	&	52	&	125	&	182	&	222	&	257	&	285\\
0.55	&	0	&	8	&	16	&	24	&	28	&	33	&	41\\
0.6	&	0	&	1	&	2	&	3	&	3	&	4	&	4\\
0.65	&	0	&	1	&	1	&	1	&	1	&	1	&	1\\
0.7	&	0	&	0	&	0	&	0	&	0	&	0	&	0\\

\hline
\end{tabular}
\label{tab:S1-2-Cs}
\end{table*}

%
%
%
\subsection{Crossing duration distribution and domain wall width sensitivity}\label{sec:sensitivity}

Before placing limits on $\Gamma_{\rm eff}$, we need to account for the fact that we do not have equal sensitivity to each domain wall width, $d$, or equivalently crossing durations, \ref{eq:tau}.
This is due in part to aspects of the clock hardware and operation, the time-resolution (data sampling frequency), and the employed search method.
Therefore, for a given $d$, we must determine the proportion of events that have crossing durations within
{${\tau_{\rm min} < \tau < \tau_{\rm max},}$}
where $\tau_{\rm min(max)}$ is the minimum (maximum) crossing duration that we are sensitive to.

\begin{figure*}[t]
\centering
	\includegraphics[width=0.4\textwidth]{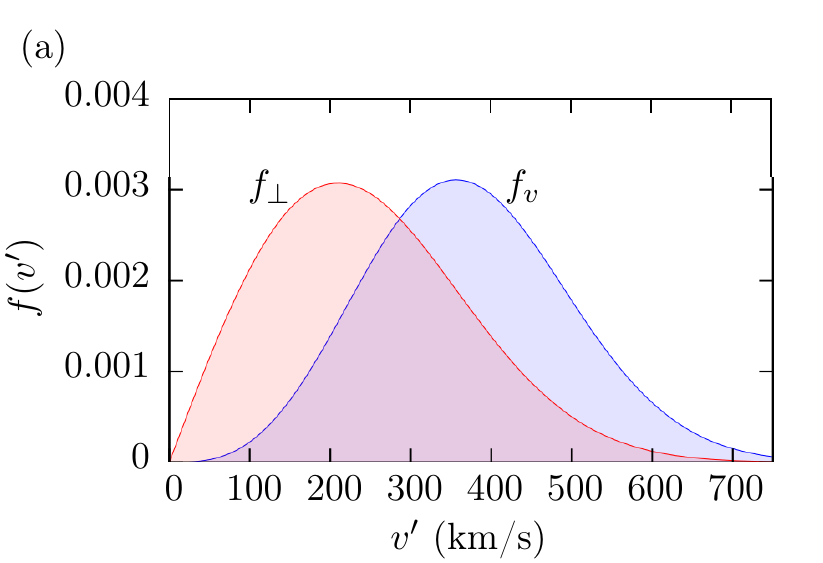}~~~~~~~~~~~
	\includegraphics[width=0.4\textwidth]{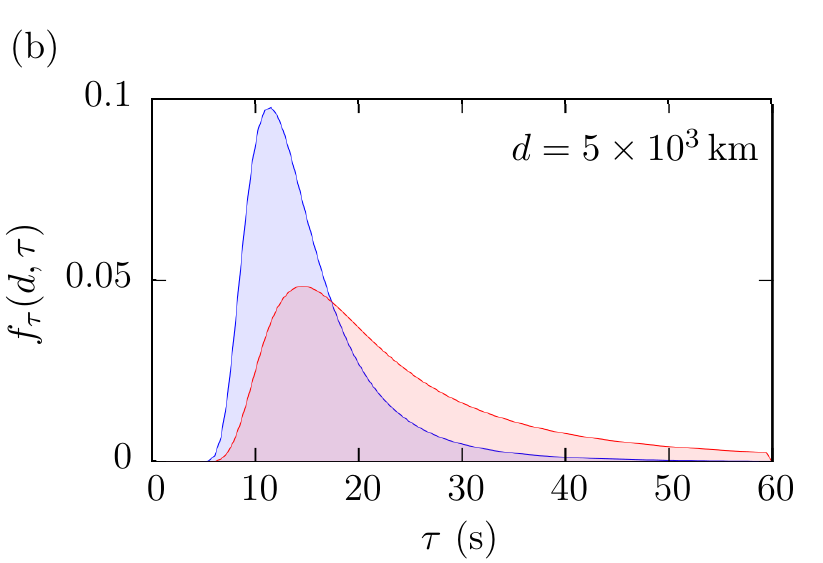}
	\caption{\small 
		{Left} panel: The blue curve shows the velocity distribution for DM objects that cross the GPS constellation (\ref{eq:distro-v}) while the red curve shows the corresponding distribution for the velocity component normal to the domain wall (\ref{eq:distro-vperp}). Right panel: the resultant single-clock crossing-time, $\tau=d/v'$, distribution for walls of width $d=5\times10^3\,$km. 
	}
\label{fig:distros}
\end{figure*}


There are two factors that determine $\tau_{\rm min}$.
The first is the so-called ``servo-loop time'' -- the fastest perturbation that can be recorded by the clock.
This servo-loop time is manually adjusted by military operators and is not known to us at a given epoch,  
however, it is known~\cite{Dupuis2008,Griggs2015} to be within $0.01$ and $0.1\,{\rm s}$.
As such, we consider the best and worse case scenarios: 
\begin{equation}
\begin{split}
\tau_{\rm min}^{\rm Servo}({\rm best)} =0.01\,{\rm s}, \\
\tau_{\rm min}^{\rm Servo}({\rm worst}) =0.1\,{\rm s}.
\end{split}
\end{equation}
Note, that below $\tau^{\rm Servo}$ we still have sensitivity to DM events, however, the sensitivity in this region is determined by response of the quartz oscillator to the temporary variation in fundamental constants.
However, the resulting limits for crossing durations shorter than the servo-loop time are generally weaker than the existing astrophysics limits (see section~\ref{sec:placinglimits}), so we consider this no further.

The second condition that affects $\tau_{\rm min}$ is the clock degeneracy: the employed GPS data set has only 30\,{\rm s} resolution, so any clocks which are affected within 30\,{\rm s} of the reference clock will not exhibit any DM-induced frequency excursion in their data (see Fig.~\ref{fig:S1} and satellites 15 and 16 in Fig.~3(a) of the main text).
For less than 60\% of the clocks to experience the jump, the (thin-wall) DM object would have to be travelling at over $700\,{\rm km\,s}^{-1}$, which is close to the galactic escape velocity (for head-on collisions), so the degeneracy does not affect the derived limits in a substantial way.
(In fact, assuming the standard halo model, less than $0.1\%$ of events are expected to have $v_\perp>700\,{\rm km\,s}^{-1}$.)
This velocity corresponds to a crossing duration for the entire network of $\sim70\,{\rm s}$.
Transforming this to crossing duration for a single clock, $\tau$, amounts to multiplying by the ratio $d/(D_{\rm GPS})$:
\begin{equation}
\tau_{\rm min}^{\rm Degen.} = \frac{d}{D_{\rm GPS}}\,70\,{\rm s}.
\end{equation}
For the thickest walls we consider ($\sim10^4\,{\rm km}$), this leads to $\tau_{\rm min}^{\rm Degen.}$ of 14\,{\rm s}.

Combining the servo-loop and degeneracy considerations, we arrive at the expression
\begin{equation}
\begin{split}
\label{eq:tmin}
\tau_{\rm min}^{\rm {best}} =
	{\rm max}\left(
		0.01 \,{\rm s} \, , \, \frac{d}{D_{\rm GPS}}\,70\,{\rm s}
	\right)  ,\\
\tau_{\rm min}^{\rm {worst}}  =
	{\rm max}\left(
		0.1 \,{\rm s} \, , \, \frac{d}{D_{\rm GPS}}\,70\,{\rm s}
	\right).
\end{split}
\end{equation}
For walls thicker than $d\sim100\,{\rm km}$, $\tau_{\rm min}$ is determined by the ${d}/(D_{\rm GPS})\,70\,{\rm s}$ term.


As to the {\em maximum} crossing duration, there are also two factors that affect $\tau_{\rm max}.$
The first, is that the wall must pass each clock in less than the sampling interval of 30\,{\rm s} -- this is the condition for the wall to be considered ``thin'':
\begin{equation}
\tau_{\rm max}^{\rm thin}=30\,{\rm s}.
\end{equation}
If a wall takes longer than 30\,s to pass by a clock, the simple single-data--point signals shown in Fig.~\ref{fig:S1} would become more complicated, and would require a more-detailed pattern-matching technique. 
Second, we only consider time ``windows'' of a certain size in our analysis (see section~\ref{sec:window}).
If a wall moves so slowly that it does not sweep all the clocks within this window, the event would be missed:
\begin{equation}
\tau_{\rm max}^{\rm window}= J_w \frac{d}{D_{\rm GPS}}.
\end{equation}

Therefore, the overall expression for $\tau_{\rm max}$ is:
\begin{equation}
\label{eq:tmax}
\tau_{\rm max}=
{\rm min}\left(
30\,{\rm s} \, , \, J_w \frac{d}{D_{\rm GPS}}
\right).
\end{equation}
Making $J_w$ large, however, also tends to increase $S^{(1)}_{\rm thresh}$ (since there is a higher chance that a large window will satisfy the condition for a ``potential event'', see section~\ref{sec:window}).
By performing the analysis for multiple values for $J_w$, we can probe the largest portion of the parameter space. 
In this work, we consider windows of $J_w$ up to 500 epochs (15000\,{\rm s}), which corresponds to a minimum velocity of $\sim4\,{\rm km\,s}^{-1}$, which is roughly the orbit speed of the satellites.
This has a negligible effect on our sensitivity, since less than $0.1\%$ of walls are expected to have $v_\perp<4\,{\rm km\,s}^{-1}$.

Assuming the standard halo model, the relative scalar velocity distribution of DM objects that cross the GPS network is quasi-Maxwellian
\begin{equation}
f_v(v)= \frac{C}{v_c^3}\,v^2\left[ \exp\left(\frac{-(v-v_c)^2}{v_c^2}\right) - \exp\left(\frac{-(v+v_c)^2}{v_c^2}\right)\right],
\label{eq:distro-v}
\end{equation}
where $v_c=220\,{\rm km\,s}^{-1}$ is the Sun's velocity in the halo frame, and $C$ is a normalisation constant. The form of Eq.~(\ref{eq:distro-v}) is a consequence of the motion of the reference frame.
However, the distribution of interest for domain walls is the perpendicular velocity distribution for walls that cross the network
\begin{equation}
f_{\perp}(v_\perp)=C'\int_{v_\perp}^\infty f_v(v) \frac{v_\perp}{v^2}\, d v,
\label{eq:distro-vperp}
\end{equation}
where $v_\perp$ is the component of the wall's velocity that is perpendicular to the wall, and $C'$ is a normalisation constant.
Note that this is not the distribution of perpendicular velocities in the galaxy -- instead, it is the distribution of perpendicular velocities that are expected to cross paths with the GPS constellation (walls with velocities close to parallel to face of the wall are less likely to encounter the GPS satellites, and objects with higher velocities more likely to).
Here, we assumed that the distribution of wall velocities is similar to the standard halo model, which is expected if the gravitational force is the main force governing wall dynamics within the galaxy. 
However, we also note that our results are fairly independent of the exact form of the velocity distribution.
Even if the actual wall velocity distribution is somewhat different, the qualitative feature of a TD ``wind'' is not expected to change.
For example, if a larger proportion of the DM objects move slower than typical galactic velocities, almost nothing changes, since then the relative velocity is essentially given by the Earth velocity in the galactic frame.

\begin{figure}
\centering
	\includegraphics[width=0.47\textwidth]{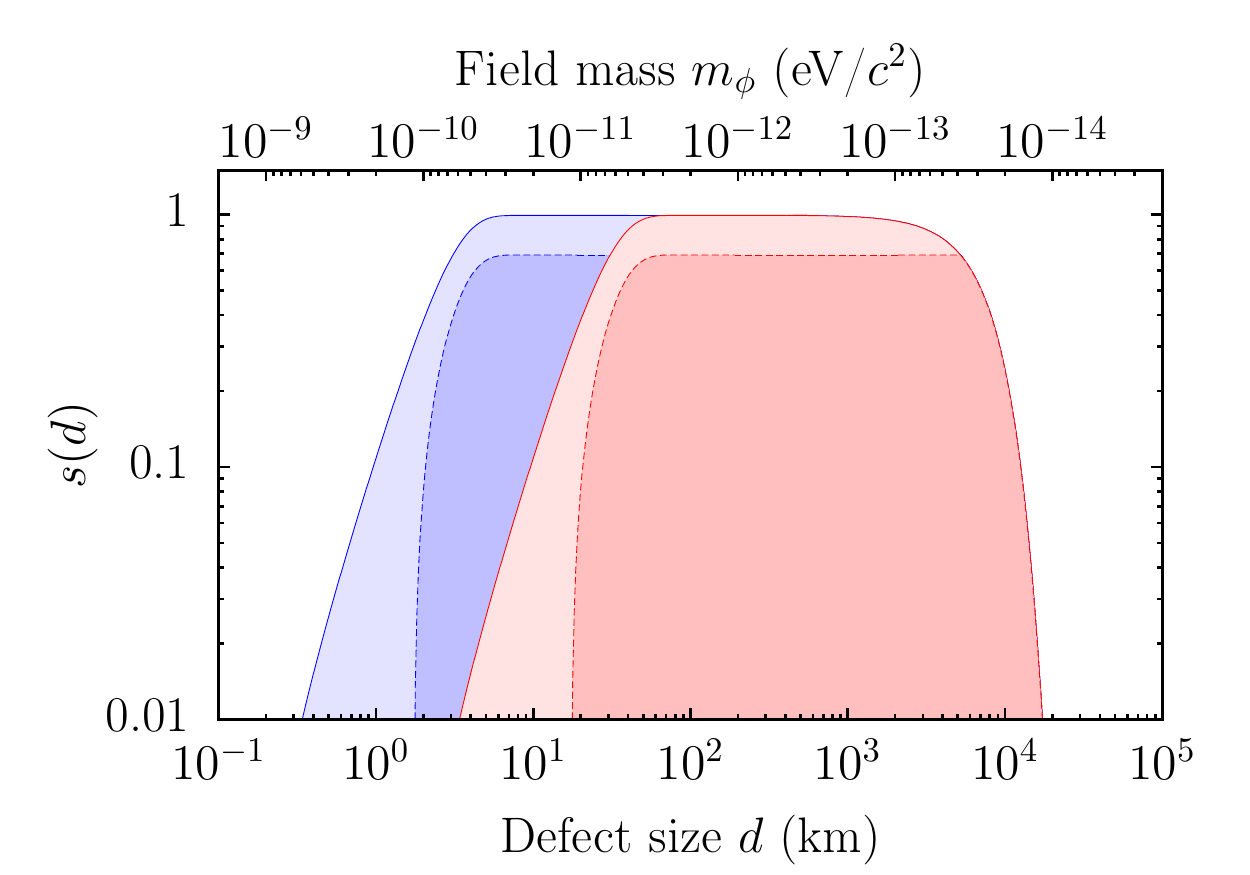}
	\caption{\small Sensitivity, $s$, as a function of the wall width $d$. The blue region corresponds to the best case scenario (due to the servo-loop uncertainty), and the red region to the worst case, see Eq.~(\ref{eq:tmin}). The darker regions correspond to a time window of size $J_w=300\,$s (i.e. $v_{\rm min}\approx170\,{\rm km\,s}^{-1}$), and the lighter regions correspond to $J_w=3000\,$s ($v_{\rm min}\approx17\,{\rm km\,s}^{-1}$).
	The upper horizontal axis shows the mass of the underlying DM field (\ref{eq:compton}).
	}
\label{fig:Sensitivity}
\end{figure}

Now, define a function $f_\tau(d,\tau)$, 
such that the integral $\int_{\tau_a}^{\tau_b}f_\tau(d,\tau)d\tau$ gives the fraction of events due to walls of width $d$ that have crossing durations between $\tau_a$ and $\tau_b$.
Note, this function must have the following normalization:
\[
\int_0^\infty f_\tau(d,\tau) \,{\rm d}\tau = 1
\]
for all $d$, and is given by
\begin{equation}
\label{distro-tau}
f_\tau(d,\tau) = \frac{d}{\tau^2}f_{\perp}(v_\perp).
\end{equation}
Plots of the velocity and crossing-time distributions are given in Fig.~\ref{fig:distros}.
Then, our sensitivity at a particular wall width is
\begin{equation}
s(d) = \int_{\tau_{\rm min}}^{\tau_{\rm max}} f_\tau(d,\tau)\, d\tau.
\end{equation}
Plots of the sensitivity function for a few various cases of parameters are presented in Fig.~\ref{fig:Sensitivity}.

\subsection{Placing limits}\label{sec:placinglimits}

The most stringent limit on the effective energy scale $\Lambda_{\rm eff}=1/\sqrt{|{\Gamma_{\rm eff}}|}$ is set by assuming one event could have occurred in the observation time, which caused a frequency excursion in the timing data that was equal in magnitude to $S^{(1)}_{\rm thresh}$.
In this case:
\begin{equation}
\mathcal{T} \to \mathcal{T}_{\rm obs},
\end{equation}
\begin{equation}
S^{(1)}_{\rm avg} \to S^{(1)}_{\rm thresh}.
\end{equation}
To determine the confidence level for our limits, we must also factor in the uncertainties.
The uncertainty from the clock noise is directly built into the $S^{(1)}$ values in Tables~\ref{tab:S1-2-Rb} and \ref{tab:S1-2-Cs}.
For our current analysis, the $S^{(1)}$ limits are substantially greater than the actual clock-noise, on the order of $5-20\sigma$.
The dominating uncertainty in our limits comes from equating the observation time with the average time between events.

To set the limits in the region where $\mathcal{T}\simeq\mathcal{T}_{\rm obs}$, we assume that the frequency of DM--GPS encounters is roughly Poissonian.
Suppose we expect to see on average $\lambda$ events in the observation time $\mathcal{T}_{\rm obs}$.
The probability for observing {\em at least} one event in the time period $\mathcal{T}_{\rm obs}$ is given by
\begin{equation}
\label{eq:Poiss}
P_{k\geq1}(\lambda)
= 1-P_0(\lambda) 
= 1-e^{-\lambda},
\end{equation}
where 
$P_{k}(\lambda)={\lambda^k e^{-\lambda}}/{k!}$
is the Poisson distribution.
For example, to place 90\% confidence level limits we require that 
$P_{k\geq1}(\lambda)=0.9.$
In this case, solving (\ref{eq:Poiss}) gives $\lambda=2.3$.
Therefore, the maximum $\mathcal{T}$ for which we can place 90\% C.L. limits is given by
\begin{equation}\label{eq:PoissFactor}
\mathcal{T}_{\rm max} = \mathcal{T}_{\rm obs}/\lambda,
\end{equation}
where for a 90\% C.L. limit, $\lambda=2.3$ and $\mathcal{T}_{\rm max}\simeq7\,{\rm yr}$.

For the region of parameter space where 
$\mathcal{T} \ll \mathcal{T}_{\rm obs}$,
our sensitivity is lower.
This is because for small values of the average time between events, the total number density of DM objects must be correspondingly high, leading to smaller DM field values inside a defect.
In the assumption that such objects constitute a significant fraction of the DM, their total energy must still add up to the total observed local DM density~\cite{Nesti2013} of 
{${\rho_{\rm DM}\approx 0.4\,{\rm GeV\,cm}^{-3}}$}:
\begin{equation}
\rho_{\rm inside} = \rho_{\rm DM} \frac{\mathcal{T} v_g}{d}.
\label{eq:rhoTD}
\end{equation}
Therefore, the energy density of each object must be smaller, and as such, the resulting signal would be smaller.


For the applicable region ($\mathcal{T}<\mathcal{T}_{\rm obs}/\lambda$), we combine the $s(d)$ function with (\ref{eq:S1}) to get the final constraints:
\begin{equation}
\label{eq:limit}
\frac{\Lambda_{\rm eff}}{d} > \sqrt{\frac{\hbar c \sqrt{\pi}\rho_{\rm DM} \, \mathcal{T}\, s(d)}{S^{(1)}_{\rm thresh}} }  .
\end{equation}
Similarly, in the case when one of the specific couplings, $\Gamma_X$, dominates over the other coupling strengths in the linear combination in Eqs.~(\ref{eq:KRb}) and (\ref{eq:KCs}), we have
\begin{equation}
\label{eq:limit-k}
\frac{\Lambda_{X}}{d} > \sqrt{\frac{\hbar c \sqrt{\pi}\rho_{\rm DM} \, \mathcal{T}\, s(d)\,K_X}{S^{(1)}_{\rm thresh}} }  .
\end{equation}
Note that the only $d$-dependence in $s(d)$ comes from the integration limits (via $\tau_{\rm min/max}$).
Expressing in convenient units:
\begin{equation}
\frac{\Lambda_{\rm eff}/{\rm TeV}}{d/{\rm km}} > 2\times10^{3} \sqrt{\frac{s(d) \, \mathcal{T}/{\rm yr}}{S^{(1)}_{\rm thresh}/{\rm ns}} } .
\end{equation}
Using the relation $d=\hbar/m_\varphi c$, one can rewrite the above limit in terms of the field mass $m_\varphi$ with the substitution
{${
(d/{\rm km} ) 
\approx {2\times10^{-10}}{} \, ({\rm eV}/m_\varphi c^2).
}$}

\begin{figure}
\centering
	\includegraphics[width=0.475\textwidth]{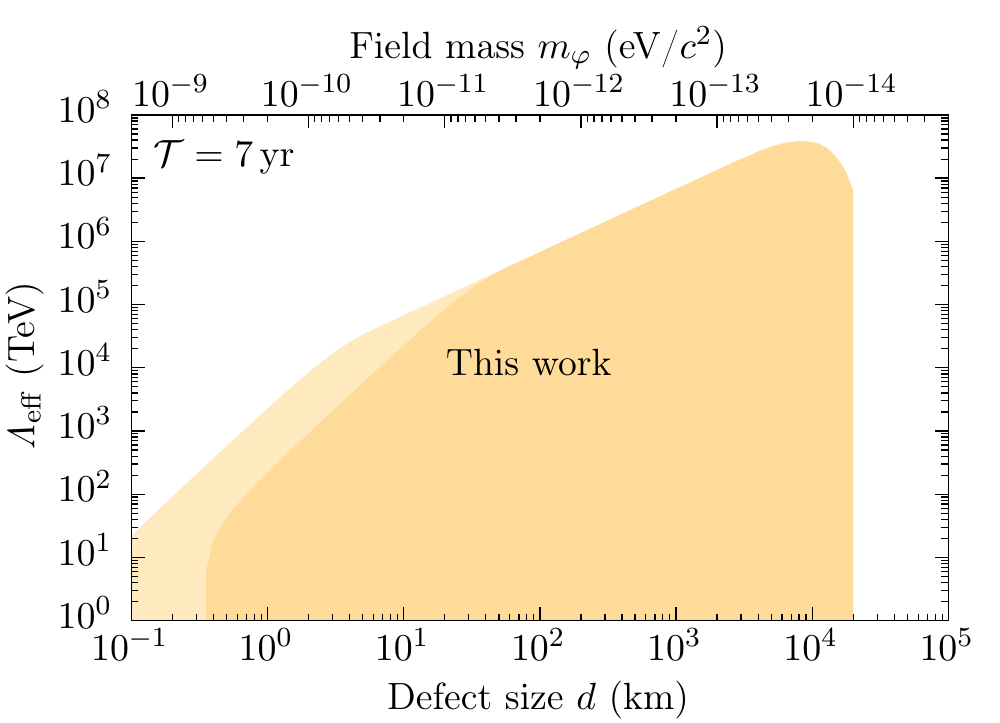}
	\caption{\small Limits (90\% C.L.) on the effective energy scale $\Lambda_{\rm eff}$ for Rb as a function of the wall width, $d$, for a fixed $\mathcal{T}=7\,{\rm yr}$. 
	The lighter and darker shaded yellow exclusion regions corresponding to the best and worst case scenarios, as described in Section~\ref{sec:sensitivity}. 
Similar (but less stringent) limits can also be placed using the Cs sub-network.
The kinks in the plot and the sharp cut off above $d\approx2\times10^4\,{\rm km}$ are due to the crossing duration sensitivities, see Eqs.~(\ref{eq:tmin}) and (\ref{eq:tmax}).
See also the contour plot in Fig.~4 of the main text.
	} 
\label{fig:limits-eff}
\end{figure}

The resulting 90\% C.L. limits on $\Lambda_{\rm eff}$ from combining the $S^{(1)}$ limits from Tables~\ref{tab:S1-2-Rb} and \ref{tab:S1-2-Cs} with Eq.~(\ref{eq:limit}), for the case when 
$\mathcal{T}=\mathcal{T}_{\rm obs}/\lambda =7\,{\rm yr} $, 
are shown in Fig.~\ref{fig:limits-eff}.
For smaller values of $\mathcal{T}$, the limits scale as $\sqrt{\mathcal{T}}$.
See also
Fig.~4  
of the main text, which shows a contour plot of the constraints as a function of $d$ and $\mathcal{T}$.

Recently, the group from Toru\'n~\cite{Wcislo2016} used an optical Sr clock to place limits on the coupling of topological defect DM to atoms.
Since this group employed an optical transition in Sr, this experiment is only sensitive to the variation in the fine structure constant (i.e., $\Lambda_\alpha$), see Eq.~(\ref{eq:KSr}).
Their data covers a period of $\mathcal{T}_{\rm obs}=45700\,{\rm s}\approx13\,{\rm hrs}$.
We combine their derived limits on $\Lambda_\alpha$ with our results in the exclusion plots. 
We note that although tight constraints can be placed using this method\cite{Wcislo2016}, in order to distinguish a true DM-induced transient event from other external sources (such as electromagnetic interference, or direct physical disturbance of the clocks), a global network is prerequisite .
One of the main advantages of our method of employing the GPS constellation is the reliance on such a global network.
Another advantage of our approach is the availability of archival data for at least the past 16 years, giving us sensitivity to the region of the parameter space with $\mathcal{T}\gtrsim1-10\,{\rm yrs}$, 
which is currently inaccessible by other methods.

If we make assumptions about the relative strengths of the couplings in  Eq.~(\ref{eq:variation}), we can place limits on individual energy scales (\ref{eq:limit-k}).
For example, in the assumption that $\Lambda_\alpha\ll\Lambda_{e/p},\Lambda_{q}$, we can place limits directly on $\Lambda_\alpha$ (and likewise for $\Lambda_{e/p}$ and $\Lambda_{q}$).
These resulting limits for $\Lambda_\alpha$ (and comparison with the results of Ref.~\citenum{Wcislo2016}) are shown in 
Fig.~5 of the main text.
Note that due to differences in the experimental technique, the optical Sr limits scale with the wall width as $d^{3/4}$, and the sensitivity of the approach of Ref.~\citenum{Wcislo2016} reduces sharply for widths greater than the Earth radius due to a frequency cut-off used in the analysis~\cite{Wcislo2016}.
In contrast, the GPS limits from this work scale linearly with $d$ (\ref{eq:limit-k}).
Both our limits and those of Ref.~\citenum{Wcislo2016} scale as $\sqrt{\mathcal{T}}$, but have sharp cut-offs above the observation time
($\mathcal{T_{\rm obs}}\sim16\,{\rm yr}$ for our work, and $\mathcal{T_{\rm obs}}\sim10^{-3}\,{\rm yr}$ for Ref.~\citenum{Wcislo2016}).
Limits on $\Lambda_{e/p}$ and $\Lambda_{q}$ (assuming these respective couplings dominate) are shown in Fig.~\ref{fig:limits-mep-q} (these couplings are unconstrained by the Sr experiment\cite{Wcislo2016}).

\begin{figure}
\centering
	\includegraphics[width=0.45\textwidth]{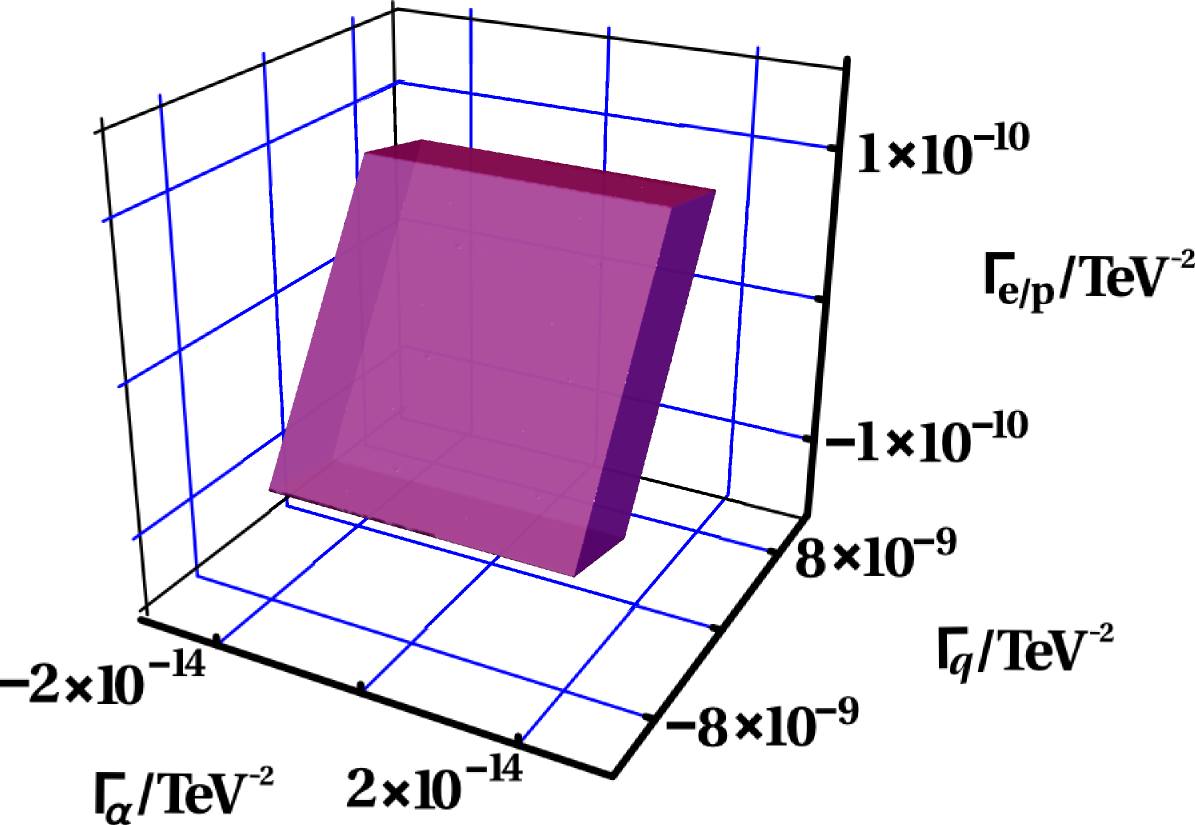}
	\caption{\small Allowed region for the coupling strength parameters $\Gamma_\alpha$,  $\Gamma_{e/p}$, and $\Gamma_{q}$, from the combined limits using the Rb and Cs GPS sub-networks, and the Sr optical clock limits from Ref.~\citenum{Wcislo2016}, for fixed $d=10^3\,{\rm km}$ and $\mathcal{T}=10^{-3}\,{\rm yr}$. Note that the allowed region (the inner part of the parallelepiped, shown in purple) is completely bound on all sides.}
\label{fig:CombineLimits3D}
\end{figure}

\begin{figure*}
\centering
	\includegraphics[width=0.475\textwidth]{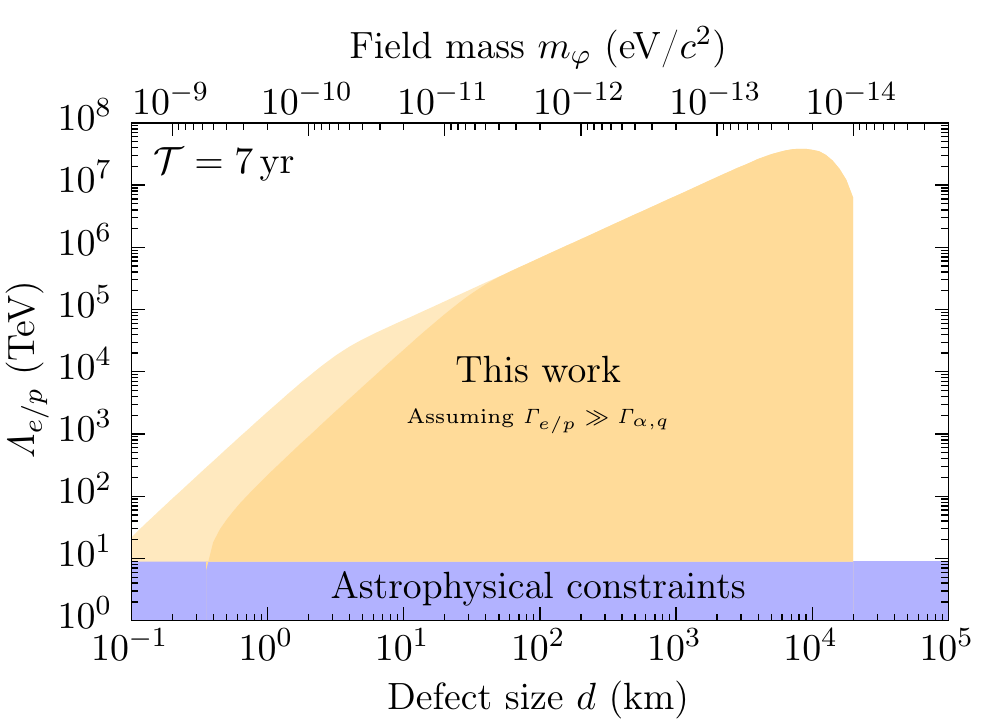}
	\includegraphics[width=0.475\textwidth]{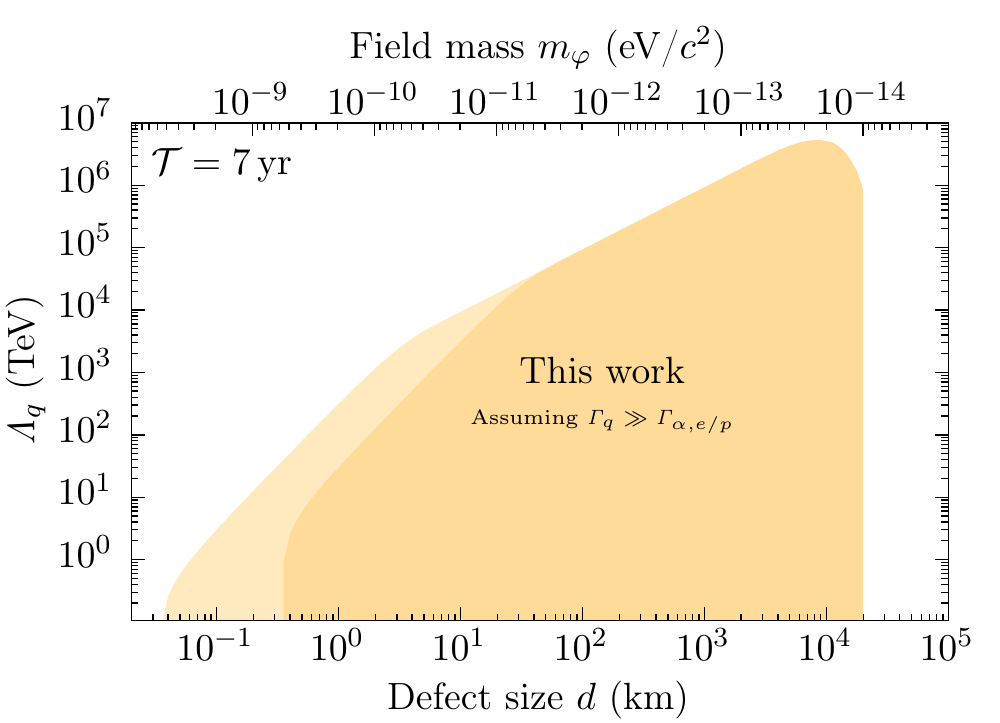}
	\caption{\small Limits (90\% C.L.) on individual energy scales from the Rb sub-network, assuming these respective couplings dominate the linear combination in Eq.~(\ref{eq:KRb}). 
Limits are shown as a function of the wall width, $d$, with fixed $\mathcal{T}=7\,{\rm yr}$, for $\Lambda_{e/p}$ (left panel) and $\Lambda_{q}$ (right panel).
These exclusion regions are shown in yellow, with the lighter and darker regions as in Fig.~\ref{fig:limits-eff}. Existing astrophysical bounds\cite{Olive2008} are shown in blue; note that $\Lambda_q$ was previously unconstrained.
	}
\label{fig:limits-mep-q}
\end{figure*}

\begin{figure*}
\centering
	\includegraphics[width=0.475\textwidth]{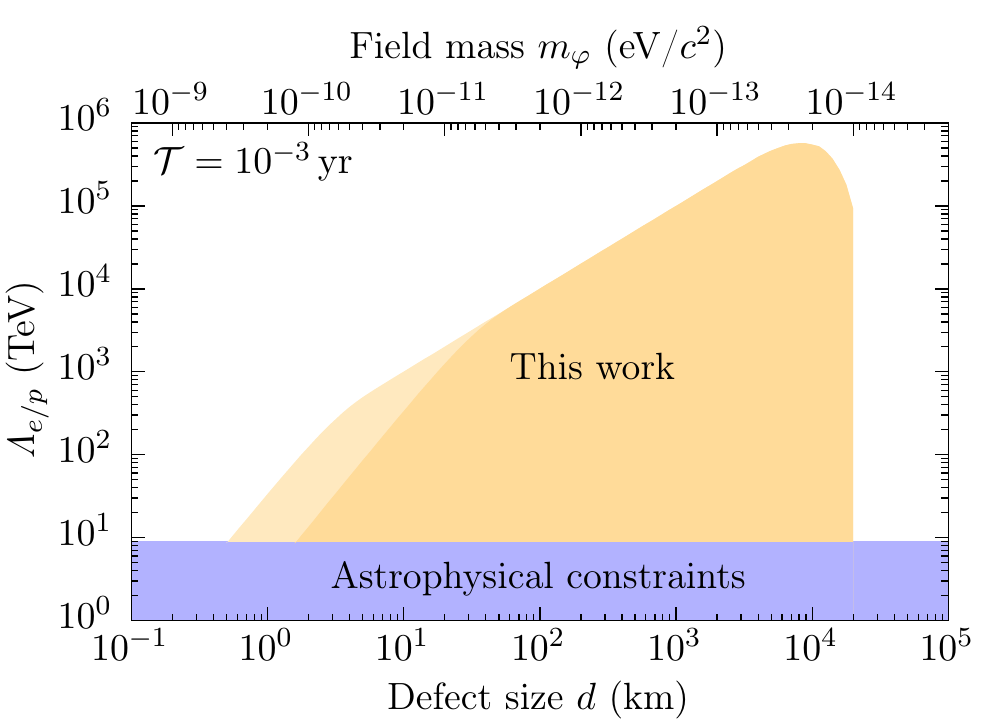}
	\includegraphics[width=0.475\textwidth]{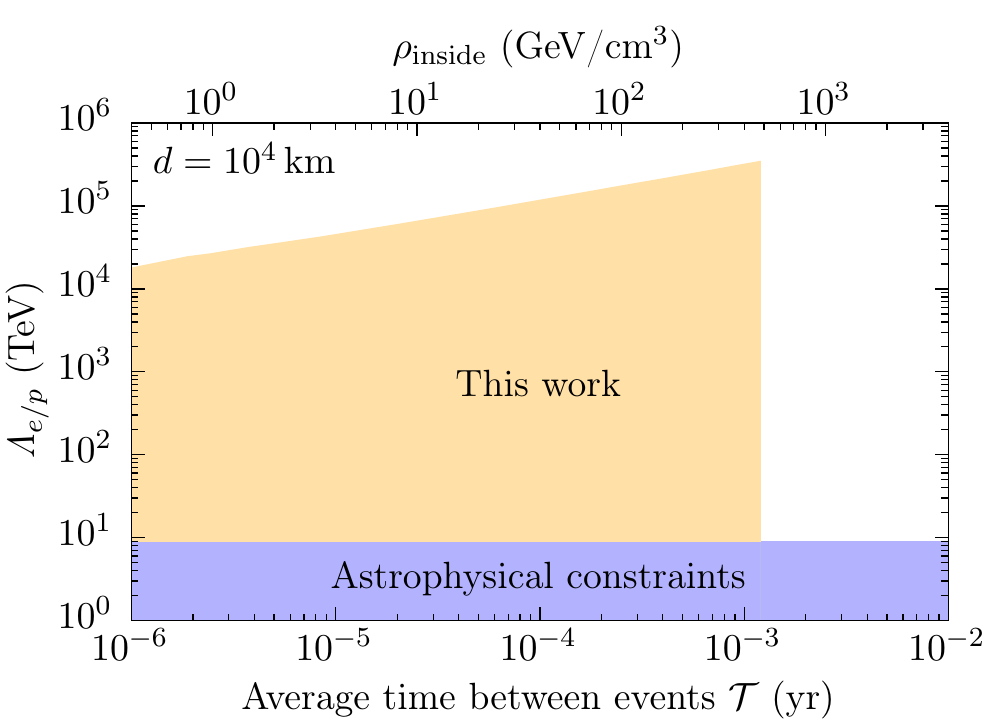}
	\caption{\small Limits on $\Lambda_{e/p}$ (90\% C.L.), by combining the limits from the Rb and Cs GPS clocks (this work) and the limits from the optical Sr clock from Ref.~\citenum{Wcislo2016}, making no assumptions on the relative coupling strengths. Left panel: as a function of $d$, with fixed $\mathcal{T}=45700\,{\rm s}$. Right panel: as a function of $\mathcal{T}$, with fixed $d=10^4\,{\rm km}$.}
\label{fig:limits-mep-comb}
\end{figure*}

\begin{figure*}
\centering
	\includegraphics[width=0.475\textwidth]{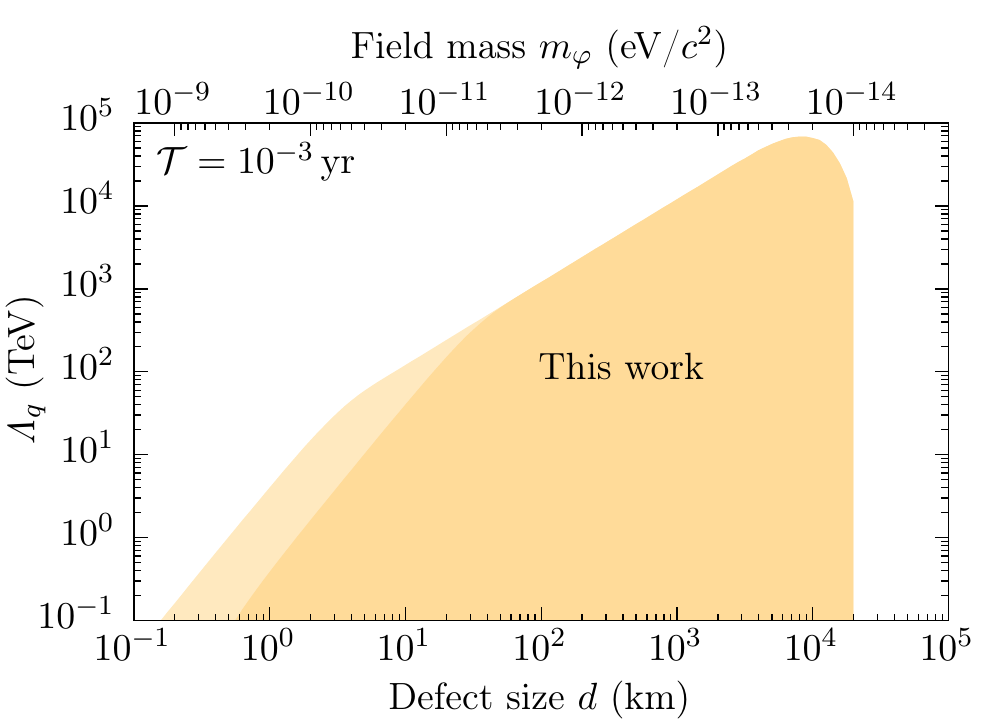}
	\includegraphics[width=0.475\textwidth]{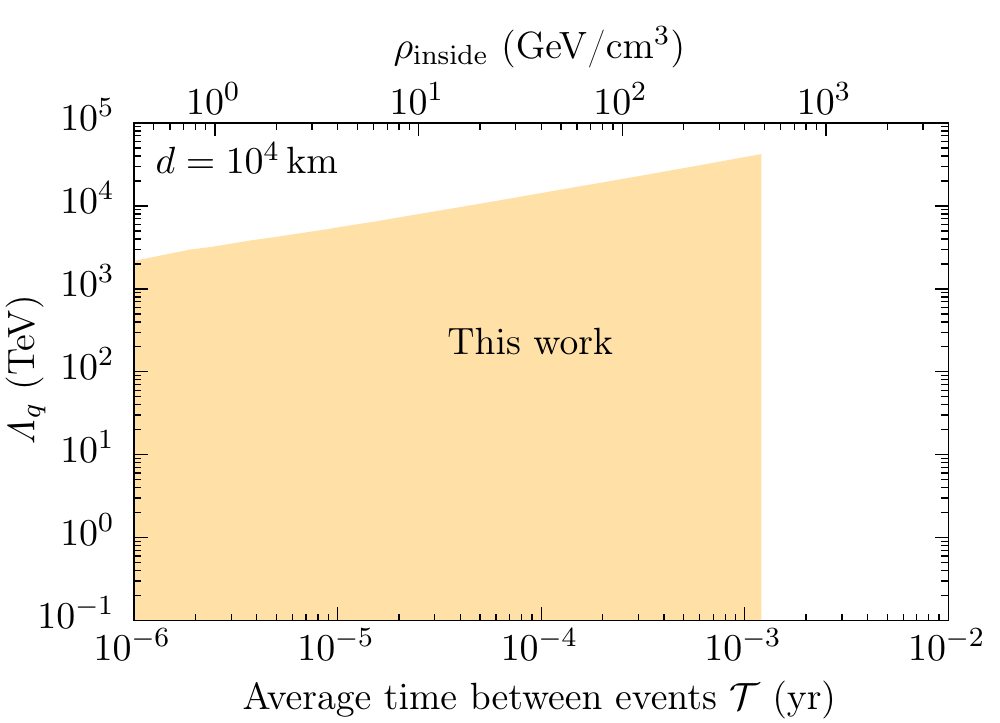}
	\caption{\small Limits on $\Lambda_{q}$ (90\% C.L.), by combining the limits from the Rb and Cs GPS clocks (this work) and the limits from the optical Sr clock from Ref.~\citenum{Wcislo2016}, making no assumptions on the relative coupling strengths. Left panel: as a function of $d$, with fixed $\mathcal{T}=45700\,{\rm s}\approx10^{-3}\,{\rm yr}$; Right panel: as a function of $\mathcal{T}$, with fixed $d=10^4\,{\rm km}$. }
\label{fig:limits-q-comb}
\end{figure*}

From the three independent limits (one from each of the Rb and Cs sub-networks determined in this work, and one from the optical Sr clock used in Ref.~\citenum{Wcislo2016}), we can derive independent limits on $\Lambda_{e/p}$ and $\Lambda_{q}$, without having to make assumptions about the relative strengths of the individual couplings.
This is possible because the three limits (from Cs, Rb, and Sr) depends on a different linear combination of the three available couplings, see Eqs.~(\ref{eq:KRb})--(\ref{eq:KSr}).
A plot showing the combined allowed region for the coupling parameters $\Gamma_\alpha$,  $\Gamma_{e/p}$, and $\Gamma_{q}$
 is presented in Fig.~\ref{fig:CombineLimits3D} 
(note $\Lambda_X=1/\sqrt{|\Gamma_X|}$).
The resulting limits on $\Lambda_{e/p}$ and $\Lambda_{q}$ are presented in Figs.~\ref{fig:limits-mep-comb} and \ref{fig:limits-q-comb}, respectively.


~\\[-2.25cm]

~\\~\\~\\
	\noindent\rule{0.01\textwidth}{0pt}
	\rule{0.05\textwidth}{0.5pt}\rule{0.35\textwidth}{1.0pt}\rule{0.05\textwidth}{0.5pt}
	\\[-1.75cm]

{
\small
\setcounter{enumiv}{77}

}


\end{document}